\documentclass[lettersize,journal]{IEEEtran}
\usepackage{amsmath,amsfonts}
\usepackage{algorithmic}
\usepackage{algorithm}
\usepackage[hidelinks]{hyperref}
\usepackage{array}
\usepackage[caption=false,font=footnotesize,labelfont=rm,textfont=rm]{subfig}
\usepackage{textcomp}
\usepackage{stfloats}
\usepackage{url}
\usepackage{verbatim}
\usepackage{graphicx}
\usepackage{cite}
\usepackage{makecell}
\usepackage{bm}
\usepackage{amssymb}
\usepackage{amsthm}
\newtheorem{thm}{Theorem} 
\newtheorem{proposition}{Proposition} 

\newcommand{\E}[1]{\mathbb{E}\left\{ #1 \right\}}
\newcommand{\tr}[1]{{\rm tr}\left( #1 \right)}
\newcommand{\norm}[1]{\left\lVert #1 \right\rVert}
\newcommand{\abs}[1]{\left\lvert #1 \right\rvert}
\newcommand{\ds}{\rm d}
\newcommand{\ps}{\rm p}
\newcommand{\Ht}{\rm H}

\newcommand{\tx}{\rm t}

\begin{document}

\title{Digital Self-Interference Cancellation in
Full-Duplex Radios: A Fundamental Limit
Perspective}

\author{}

\author{Limin Liao, Jun Sun, Junzhi Wang, Chao Deng, Jizhao Wang, Fangsen Li and Yingzhuang Liu
\thanks{Limin Liao, Jun Sun, Junzhi Wang, Chao Deng, Yingzhuang Liu are with the School of Electronic Information and Communications, Huazhong University of Science and Technology, Wuhan, China. (E-mail: liaolimin001@hust.edu.cn; juns@hust.edu.cn; wangjunzhi@hust.edu.cn; d201980612@hust.edu.cn; liuyz@hust.edu.cn)}
\thanks{
Jizhao Wang and Fangsen Li are with the China Mobile Group Device Co., Ltd. (E-mail: wangjizhao@cmdc.chinamobile.com; lifangsen@cmdc.chinamobile.com)}
}

\markboth{IEEE Transactions on Communications}%
{Shell \MakeLowercase{\textit{et al.}}: A Sample Article Using IEEEtran.cls for IEEE Journals}


\maketitle

\begin{abstract}

Digital self-interference cancellation (D-SIC) is of crucial importance for the implementation of in-band full-duplex (IBFD) radios. 
Unfortunately, the achievable performance limit remains underexplored.
To fill this gap, in this paper we aim to explore the performance limit, i.e., the minimum residual self-interference (RSI) of the most commonly used parallel Hammerstein (PH) canceller, and provide the achievable pilot design accordingly. To this end, we first conduct a \textit{systematic} analysis of the RSI power for the PH canceller, which takes into account both the \textit{truncation}-induced error and the \textit{noise}-induced error, whereas the former is usually ignored in the existing works. To simplify the performance analysis of RSI power, we employ the generalized Laguerre polynomial (GLP)-based PH canceller instead of the conventional monomial-based one, due to the appealing orthogonality property of the GLP for Gaussian inputs. With the GLP representation of the PH canceller, we further prove that the least-squares channel estimator is \textit{asymptotically} \textit{unbiased}, thus demonstrating the asymptotic optimality of Gaussian pilot sequences. Moreover, for the pilot sequence with a finite length, a succinct criterion for minimizing the RSI, namely, the condition-number-to-minimum-eigenvalue ratio (CMER) criterion, which essentially balances the truncation-induced and noise-induced error, is presented. By contrast, the existing works normally consider the latter only. Interestingly, it is revealed that an appropriate peak-to-average power ratio (PAPR) of the pilot sequence is of critical importance to achieve the above balance. Simulation results demonstrate that the pilot sequence optimized according to our proposed CMER criterion can achieve an RSI as low as -87.3 dBm, which is over 14 dB lower than that of the conventional pilot sequence (HE-LTF) and over 6 dB lower than that of the state-of-the-art pilot sequence proposed in \cite{Brink24}, provided that the order of the PH canceller is no higher than 9 because of the complexity constraint.

\end{abstract}


\section{Introduction}
\IEEEPARstart{I}{n}-band full-duplex (IBFD) radios have great potential to double the spectral efficiency of wireless systems by allowing simultaneous transmission and reception over the same frequency band \cite{mohammadi2023comprehensive}. 
The key challenge in implementing IBFD lies in the fact that the extremely strong self-interference compared to the noise floor (for example, more than 110 dB higher) should be suppressed to the noise floor level or below.

To achieve the above challenging goal, we normally need to perform the self-interference cancellation (SIC) in both the analog and digital domains, corresponding to analog SIC (A-SIC) and digital SIC (D-SIC), respectively \cite{bharadia2013full,korpi2016full,smida2023full}. As the first stage of SIC, A-SIC is required to reduce the self-interference (SI) to a sufficiently low level---within the dynamic range of the analog-to-digital converter (ADC)---through radio-frequency analog cancellation circuits and antenna isolation. However, because of hardware limitations, the performance of A-SIC remains limited. Actually, from the perspective of fundamental limits, it is demonstrated in \cite{liao2025analog} that it is rather difficult to achieve more than 60 dB of SIC in the analog domain.
Consequently, the residual self-interference (RSI) must be addressed in the digital domain (i.e., by D-SIC).


The key challenge of D-SIC is to eliminate the nonlinear distortions in the SI, which arise mainly from the power-amplifier (PA) nonlinearity and in-phase/quadrature (I/Q) imbalance. There have been many works in this line \cite{ahmed2015all,korpi2014full,bharadia2013full,anttila2013cancellation,anttila2014modeling,islam2019comprehensive,elsayed2020low,korpi2015adaptive,komatsu2020iterative}. \cite{bharadia2013full} models the nonlinear distortions of the transmitter through memory polynomials, and 
was the first to implement an end-to-end full-duplex Wi-Fi prototype, verifying the feasibility of 110 dB cancellation performance. The parallel Hammerstein (PH) canceller was proposed to capture the nonlinear distortions of transmitter (Tx) chains and the memory effect of the SI channel \cite{anttila2013cancellation}. \cite{anttila2014modeling} incorporates I/Q imbalance into the PH canceller and extends it to multiple-input multiple-output scenarios. Considering all nonlinearities of the Tx and receiver (Rx) chains, \cite{islam2019comprehensive} provides a digital canceller based on the Volterra-series model. A nonlinear D-SIC based on deep neural networks has been proposed in \cite{elsayed2020low}, which achieves performance comparable to polynomial models with reduced complexity. However, its ability to track channels in real time remains questionable. Moreover, the orthogonalization-based least-mean-square (LMS) algorithms were proposed in \cite{korpi2015adaptive,korpi2016full,li2018augmented} for tracking time-varying channels. To further reduce complexity, \cite{komatsu2020iterative,ahmed2013self} proposed iterative parameter estimation methods to decouple the estimation of the SI channel responses and nonlinear distortions.

Despite these advances, the achievable RSI limit of D-SIC with a finite order remains underexplored. As a companion to \cite{liao2025analog}, which addresses the fundamental limit of A-SIC, in this paper, we aim to close this gap by exploring the performance limit of D-SIC from a fundamental perspective, i.e., how small the RSI can be under a given model order, and how it can be achieved in practice. To address these problems, we will focus on the PH canceller with the least-squares (LS) estimator. The reasons for employing the LS estimator (rather than the LMS estimator) are as follows: 1) The LS estimator has higher convergence speed and relatively lower computational complexity than LMS, especially in fast-varying channels; 2) Due to the adaptive and iterative nature of LMS, it is more difficult for LMS to analyze the RSI and design the optimal pilot sequence accordingly. Meanwhile, we mainly consider the PA-induced nonlinearity without considering the other nonlinear effects such as I/Q imbalance and low-noise-amplifier (LNA) nonlinearity for the reasons as follows: 1) I/Q imbalance normally varies much more slowly than the SI channel, therefore it is quite reasonable to assume high-accuracy I/Q calibration and thus the I/Q imbalance could be ignored; 2) RSI usually falls within the linear region of the LNA since the A-SIC is responsible for ensuring RSI within the dynamic range of ADCs.

To explore the performance limit of the PH canceller with LS estimation, the core tasks are to characterize the RSI in a systematic manner and to design the optimal pilot sequence according to some succinct criteria that aim to minimize the RSI in an exact or approximate sense. It is worth noting that the optimal pilot sequence design in D-SIC bears some similarity to the pilot design in channel estimation; the key distinction between them is: the former cares about the \textit{reconstruction error} of SI, while the latter only cares about the \textit{estimation error} of the channel itself. If we decompose the reconstruction error of SI, it is easy to find that the estimation error is just one component of the former, i.e., the \textit{noise}-induced reconstruction error, whereas another key component is the \textit{bias}-induced reconstruction error (due to truncation), which exhibits an opposite trend to the noise-induced one in the case of a high peak-to-average power ratio (PAPR). Thus, a delicate balance between them (i.e., the bias--variance trade-off) is of crucial importance for achieving the best performance. In this regard, Komatsu \textit{et al.} developed a theoretical framework for
PH cancellers using generalized Laguerre polynomials (GLP) and analyzed the residual SI power \cite{komatsu2021theoretical}. However, their analysis does not consider the reconstruction error induced by the estimation method and address the corresponding pilot-design problem. Similarly, \cite{Brink24} explored the pilot design problem in the GLP analytical framework; however, it ignored the bias-induced reconstruction error and only considered the noise-induced one, thus potentially incurring significant performance loss. In fact, it turns out its performance loss could be more than 6 dB as compared with our proposed pilot design scheme seeking the bias--variance trade-off as described in the remainder of this paper.

The contributions of this paper are summarized as follows:
\begin{itemize}
  \item We provide a \textit{systematic} analysis of the RSI for finite-order LS-based PH cancellers, which jointly accounts for \textit{noise}-induced and \textit{bias}-induced reconstruction errors, whereas existing works mainly consider the former and ignore the latter.

  \item 
  We prove the asymptotic \textit{unbiasedness} of LS estimation with Gaussian pilot sequences by taking advantage of the GLP representation of the PH canceller, thus demonstrating the asymptotic optimality of Gaussian sequences and providing a valuable benchmark for finite-length pilot design.


  \item We reveal that PAPR plays a critical role in both the noise-induced reconstruction error (NIRE, which favors high PAPR) and the bias-induced reconstruction error (BIRE, which favors low PAPR). Consequently, we propose a succinct criterion, \textit{condition-number-to-minimum-eigenvalue ratio} (CMER), to choose the pilot sequence that approaches the best BIRE--NIRE trade-off. It turns out that our scheme based on this criterion can achieve a gain of more than 6 dB over the state-of-the-art scheme in \cite{Brink24} if the PH order is constrained to be no more than 9.  
\end{itemize}

The rest of this paper is organized as follows: In Section \uppercase\expandafter{\romannumeral2}, we provide the relevant theoretical background and present the mathematical models for PH cancellers. In Section \uppercase\expandafter{\romannumeral3}, we detail the analysis of the RSI of the PH canceller under LS estimation, and propose the optimization criterion for pilot sequences. Additionally, we summarize various factors that affect the performance of the PH canceller. In Section \uppercase\expandafter{\romannumeral4}, we validate the optimization criterion by comparing the performance under different pilot sequences through numerical simulations, and evaluate the influence of several factors. Finally, the conclusion is provided in Section \uppercase\expandafter{\romannumeral5}. 

\emph{Notation:} In this paper, the circularly symmetric complex Gaussian variable with variance $\sigma^2$ is denoted as $\mathcal{CN}(0,\sigma^2)$, and an exponential distribution is denoted as $\exp(\lambda)$. The operator $*$ represents the convolution, and $\otimes$ represents the Kronecker product. $z^*$ is the conjugate of the complex $z$, and $|z|$ is the modulus of $z$. For a matrix $\boldsymbol{X}$, $\boldsymbol{X}^\mathrm{T}$ is the transpose of $\boldsymbol{X}$, $\boldsymbol{X}^\mathrm{H}$ is the conjugate transpose of $\boldsymbol{X}$, $\boldsymbol{X}^{-1}$ is the inverse of $\boldsymbol{X}$, $\mathrm{tr}(\boldsymbol{X})$ represents the trace of $\boldsymbol{X}$, $\|\boldsymbol{X}\|_\mathrm{F}$ represents the Frobenius norm of $\boldsymbol{X}$, and $\|\boldsymbol{X}\|$ represents the 2-norm of $\boldsymbol{X}$. Particularly, we denote the diagonal matrix $\boldsymbol{A}=[a_{ij}]\in \mathrm{C}^{N\times N}$ by $\mathrm{diag}(a_{11}\ \dots\ a_{NN})$.

\section{Background and Modeling}
In this section, we first provide the model of the classical PH canceller for D-SIC. For the sake of analysis convenience, the conventional monomial-based PH canceller is equivalently expressed as the Laguerre Polynomial-based one, while the latter has the advantage of mutual orthogonality among the basis functions.

\subsection{Modeling of PH Cancellers}
\begin{figure}[!t]
\centering
\includegraphics[width=3.2in]{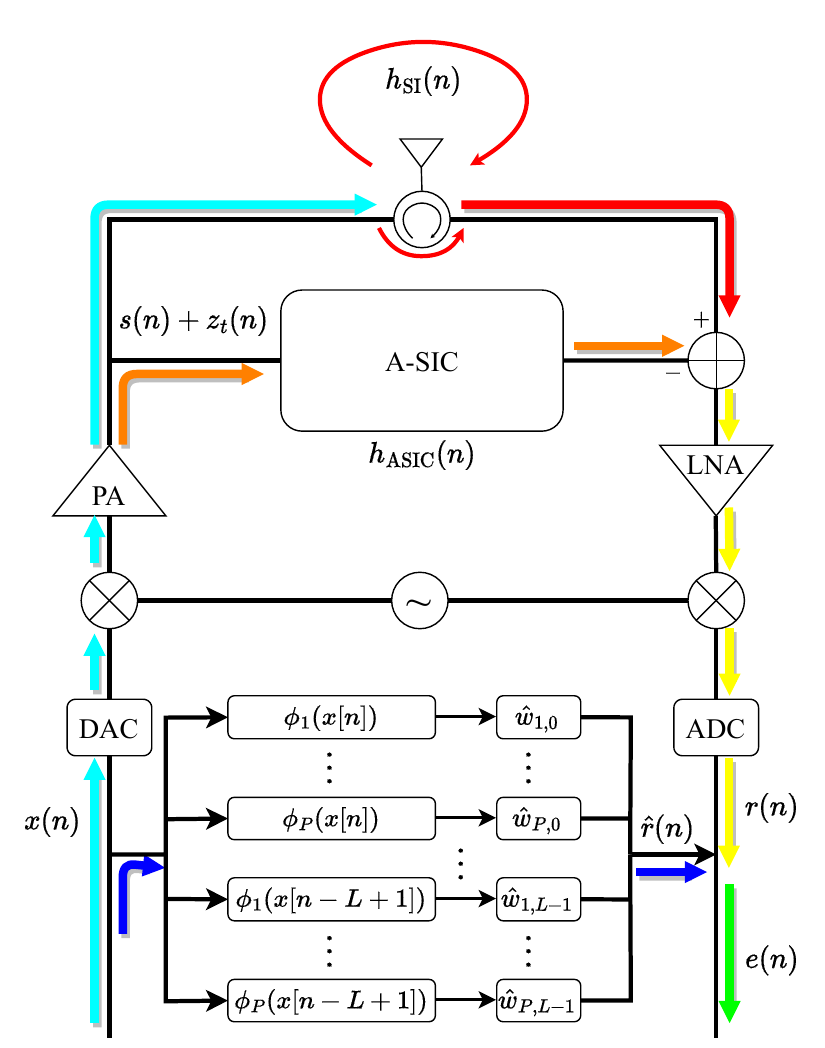}
\caption{Digital self-interference cancellation based on the PH canceller.}
\label{PH_canceller}
\end{figure}
As depicted in Fig. \ref{PH_canceller}, taking into account the nonlinear distortions at the Tx chain, the actual Tx symbols $s[n]$ can be expressed as
\begin{align}\label{s[n]}
    s[n] = \sum_{\substack{p=1\\p = {\rm odd}}}^{+\infty}c_p\big|x[n]\big|^{p-1}x[n],
\end{align}
where $c_p$ is the coefficient of the $p$th-order nonlinear component and $x[n]$ is the modulated symbol. For the sake of simplicity, we define the $p$th-order basis function as
\begin{align}\label{phi_p}
    \phi_p\big(x[n]\big) =  \big|x[n]\big|^{p-1}x[n].
\end{align}
We assume that A-SIC can effectively suppress SI so that the RSI is within the dynamic range of the ADC and the LNA behaves as a linear amplifier. Consequently, by taking the Tx noise $z_\mathrm{t}[n]$ into account, the SI at the Rx chain after A-SIC can be expressed as
\begin{align}\label{r[n]}
    r[n] = h_\mathrm{res}[n]*\big(s[n]+z_\mathrm{t}[n]\big)+z_\mathrm{r}[n],
\end{align}
 where $z_\mathrm{r}[n]$ is the Rx noise that includes the thermal noise and the quantization noise, and the equivalent channel $h_\mathrm{res}[n]$ can be expressed as
\begin{align}
    h_\mathrm{res}[n] = &h_\mathrm{SI}[n]-h_\mathrm{ASIC}[n]\notag\\
    = &\sum\limits_{\ell=0}^{L-1} \alpha_{\ell}\delta[n-\ell].
\end{align}
Here, $h_\mathrm{SI}[n]$ is the SI channel response that includes spatial channels and the memory of nonlinear components, $h_\mathrm{ASIC}[n]$ is the channel response of A-SIC, and $L$ is the number of channel taps. Therefore, the Rx SI in Eq. \eqref{r[n]} can be expressed as
\begin{align}\label{r[n]_PH}
    r[n] = &\sum\limits_{\substack{p=1\\p={\rm odd}}}^{\infty}\sum\limits_{\ell=0}^{L-1} w_{p,\ell}\phi_p\big(x[n-\ell]\big)+z[n],
\end{align}
where $z[n]\sim\mathcal{CN}(0,\sigma_z^2)$ is the equivalent Rx noise, which can be expressed as
\begin{align}
     z[n] = h_\mathrm{res}[n]*z_\mathrm{t}[n]+z_\mathrm{r}[n].
\end{align}
Similarly, for an $P$-order PH canceller with memory length $L$ , the reconstructed signal can be expressed as
\begin{align}\label{PH_Eq}    
\hat{r}[n] = &\sum\limits_{\substack{p=1\\p={\rm odd}}}^{P}\sum\limits_{\ell=0}^{L-1} \hat{w}_{p,\ell}\phi_p\big(x[n-\ell]\big).
\end{align}
To simplify the notation, defining the number of weights as $L_w=L(P+1)/2$, the reconstructed signal can be rewritten in the matrix-vector form as follows:
\begin{align}\label{PH_model}
    \hat{r}[n] = \boldsymbol{\phi}^\mathrm{T}\big(x[n]\big)\hat{\boldsymbol{w}},
\end{align}
where the regression vector $\boldsymbol{\phi}\big(x[n]\big)\in \mathbb{C}^{L_w}$ is defined as
\begin{align}
    \boldsymbol{\phi}^\mathrm{T}\big(x[n]\big)=\Big[\boldsymbol{\phi}^\mathrm{T}_{P}\big(x[n]\big)\ \cdots\ \boldsymbol{\phi}_{P}^\mathrm{T}\big(x[n-L+1]\big)\Big]
\end{align}
in which the nonlinear basis vector $\boldsymbol{\phi}_{P}\in\mathbb{C}^{(P+1)/2}$ is defined as
\begin{align}
    \boldsymbol{\phi}_{P}^\mathrm{T}\big(x[n]\big) = \Big[\phi_1\big(x[n]\big)\ \phi_3\big(x[n]\big)\ \cdots\ \phi_P\big(x[n]\big)\Big],
\end{align}
and the weight vector $\hat{\boldsymbol{w}}\in\mathbb{C}^{L_w}$ is
\begin{align}\label{Vec_w}
    \hat{\boldsymbol{w}}&=[\hat{w}_{1,0}\ \cdots\ \hat{w}_{P,0}\ \cdots\ \hat{w}_{1,L-1}\ \cdots\ \hat{w}_{P,L-1}]^\mathrm{T}.
\end{align}
Similarly, the Rx SI can be rewritten as
\begin{align}\label{r[n]_phi}
    r[n] = \boldsymbol{\phi}^\mathrm{T}\big(x[n]\big)\boldsymbol{w}+\varepsilon\big(x[n]\big)+z[n],
\end{align}
where $\varepsilon\big(x[n]\big)$ is the truncation error defined as 
\begin{align}
    \varepsilon\big(x[n]\big)=\sum\limits_{\substack{p>P\\p={\rm odd}}}^{\infty}\sum\limits_{\ell=0}^{L-1} w_{p,\ell}\phi_p\big(x[n-\ell]\big).
\end{align}
Therefore, the RSI can be expressed as
\begin{align}\label{RSI_phi}
    e[n] &= r[n]-\hat{r}[n]  \notag \\
    &=\underbrace{\boldsymbol{\phi}^\mathrm{T}\big(x[n]\big)(\boldsymbol{w}-\hat{\boldsymbol{w}})}_{\text{Reconstruction error}}
    +\underbrace{\varepsilon\big(x[n]\big)}_{\text{Truncation error} }+\underbrace{z[n]}_{\text{Noise}},
\end{align}
which can be decomposed as:
\begin{itemize}
    \item \textit{Truncation error:} the truncation error caused by insufficient order of the PH canceller determines the theoretical upper bound of the cancellation performance of the PH canceller.
    \item \textit{Reconstruction error:} the reconstruction error is caused by the estimation error of the weights due to the noise and the truncation error, determining the actual performance of the PH canceller.
    \item  \textit{Noise:} The equivalent Rx noise, including the Tx noise, the Rx thermal noise, and the quantization noise, is generally dominated by the Rx thermal noise since the other noise components need to be sufficiently suppressed by A-SIC.
\end{itemize}

Normally, the dominant term in the RSI is the reconstruction error, since one can choose an appropriate order of the PH canceller such that the truncation error is below the noise level. To facilitate the analysis of the reconstruction error, it is quite desirable to employ an orthogonal expansion model for the PH canceller. However, the conventional monomial-based (i.e., $\phi_p\big(x[n]\big)$) PH is a non-orthogonal one. Fortunately, when the input is Gaussian, such an orthogonal expansion indeed exists, namely the generalized Laguerre polynomial expansion.





\subsection{Orthogonal Expansion of PH Canceller via Generalized Laguerre Polynomials (GLP)} \label{GLP}
The conventional PH canceller is expanded by the monomial basis functions, i.e., $ \big|x[n]\big|^{p-1}x[n]$, which are non-orthogonal in nature. To facilitate the analysis of RSI, we resort to the orthogonal expansion of the PH canceller, which is made possible by the Laguerre polynomials.

The generalized Laguerre polynomials with parameter 1 are the orthogonal polynomials with respect to the exponential weights (or distribution). Mathematically, GLP is defined as \cite{komatsu2021theoretical} 
\begin{align}
    L_k^1(u) = \sum_{r=0}^{k} \frac{(-1)^r}{r!}\binom{k+1}{r+1}u^{r},
\end{align}
and we have 
\begin{align}\label{ALP}
    \mathbb{E}\big\{L_r^1(u)L_k^1(u)u\big\}&=\int_0^\infty L_r^1(u)L_k^1(u)ue^{-u}{\ds}u\notag\\
    &=(k+1)\delta_{rk},
\end{align}
if $u$ is exponentially distributed.


When the Gaussian input is independent and identically distributed (i.i.d.), i.e., $x[n]\overset{i.i.d.}{\sim}\mathcal{CN}(0,1)$, its squared modulus $\big|x[n]\big|^2$ is exponentially distributed, thus we have 
\begin{align}\label{op_GLP}
    &\mathbb{E}\Big\{L^1_{r}\big(\big|x[n]\big|^2\big)x[n]L_k^1\big(\big|x[m]\big|^2\big)x^*[m]\Big\}\notag\\
    =&(r+1)\delta_{rk}\delta_{nm},
\end{align}
where $\delta_{rk}$ and $\delta_{mn}$ denote the Kronecker delta function.

Thus, the polynomials $L^1_{r}\big(\big|x[n]\big|^2\big)x[n]$ constitute a complete orthogonal set and therefore provide an ideal choice for the expansion of the PH canceller. Specifically, we define the $q$th-order ($q=2k+1$) basis function based on Laguerre polynomial as 
\begin{align}
    \psi_{q}\big(x[n]\big) &= \frac{1}{\sqrt{k+1}}L^1_{k}\big(\big|x[n]\big|^2\big)x[n]\notag\\
    &=\sum_{r=0}^{k} l_{2k+1,2r+1}\big|x[n]\big|^{2r}x[n]\notag\\
    &=\sum_{\substack{p=1\\p={\rm odd}}}^{q} l_{q,p}\phi_{p}\big(x[n]\big),
\end{align}
where
\begin{align}
    l_{2k+1,2r+1}=\frac{1}{\sqrt{k+1}}\frac{(-1)^{r}}{r!}\binom{k+1}{r+1},\
     k\ge r\ge 0,
\end{align}
and we have
\begin{align}\label{op_GLP}
    \mathbb{E}\Big\{\psi_q\big(x[n]\big) \cdot \psi_p^*\big(x[m]\big)\Big\} =\delta_{qp}\delta_{nm}.
\end{align}
Therefore, for $x[n]\overset{i.i.d.}{\sim}\mathcal{CN}(0,1)$, the Rx SI in Eq. \eqref{r[n]_phi} can be orthogonally expanded as
\begin{align}\label{r[n]_GLP}
    r[n] =\boldsymbol{\psi}^\mathrm{T}\big(x[n]\big)\boldsymbol{\theta}
    +\epsilon\big(x[n]\big)
    +z[n],
\end{align}
where the regression vector $\boldsymbol{\psi}\big(x[n]\big)$ is obtained by replacing the basis functions in $\boldsymbol{\phi}\big(x[n]\big)$ from $\phi_p(\cdot)$ to $\psi_p(\cdot)$ for $p=1,3,\dots,P$, the weights $\theta_{q,\ell}$ are related to $w_{p,\ell}$ by
\begin{align}
    w_{p,\ell} = \sum_{\substack{q=p\\q = {\rm odd}}}^\infty \theta_{q,\ell}\cdot l_{q,p},
\end{align}
and the truncation error can be expressed as
\begin{align}
    \epsilon\big(x[n]\big) = \sum\limits_{\substack{q>P\\q={\rm odd}}}^{\infty}\sum\limits_{\ell=0}^{L-1} \theta_{q,\ell}\psi_q\big(x[n-\ell]\big).
\end{align}
Similarly, the reconstructed signal can be rewritten as
\begin{align}\label{PH2GLP}
    \hat{r}[n] &= \boldsymbol{\phi}^\mathrm{T}\big(x[n]\big)\hat{\boldsymbol{w}}\notag\\
    &=\boldsymbol{\psi}^{\rm{T}}\big(x[n]\big)\boldsymbol{T}^{-1}\hat{\boldsymbol{w}},
\end{align}
where the transform matrix $\boldsymbol{T}\in\mathbb{C}^{L_w\times L_w}$ can be expressed as
\begin{align}
    \boldsymbol{T} = \boldsymbol{I}_{L}\otimes\boldsymbol{L}_P
\end{align}
in which 
\begin{equation}
\boldsymbol{L}_P = 
\begin{bmatrix} l_{1,1} & l_{3,1} & \cdots & l_{P,1} \\ 
0 & l_{3,3} & \cdots & l_{P,3}\\
\vdots & \vdots & \ddots & \vdots\\
0 & 0 & \cdots & l_{P,P}\end{bmatrix}.
\end{equation} 
It is worth noting that this transformation lays the foundation for the equivalence of the GLP and PH representations under LS estimation. Compared to the conventional PH canceller, the GLP representation makes the analysis of RSI more tractable. More importantly, we can establish the asymptotic unbiasedness of LS estimation of the weight $\boldsymbol{\theta}$
by exploiting the orthogonality in Eq. \eqref{op_GLP}, which proves to play a critical role for choosing the optimal pilot sequence.

\section{Theoretical analysis and Pilot optimization}
In this section, we develop a tractable RSI analysis framework for the LS-based PH canceller. By exploiting the equivalence between the PH and GLP representations under the LS estimator, the RSI can be decomposed into the truncation error, the bias-induced reconstruction error (BIRE), the noise-induced reconstruction error (NIRE), and the Rx noise. Next, we prove the asymptotic unbiasedness of LS estimation based on Gaussian sequences. Then we analyze the BIRE--NIRE trade-off in the non-asymptotic case, which is proven to be strongly dependent on the PAPR of pilot sequences, and propose a robust criterion for pilot sequence design. Finally, we summarize the key factors that affect the performance of PH cancellers.

\subsection{Analysis of the RSI}
\subsubsection{ PH cancellers}
\ \newline
\indent According to the Rx SI in Eq. \eqref{r[n]_phi}, for a sequence vector $\boldsymbol{x}_{\circ}\in\mathbb{C}^{L_{\circ}},\ (L_{\circ}\ge L)$, where $\circ\in\{\rm{p},\rm{d}\}$ represents the \textit{pilot} and \textit{data}, respectively, the corresponding Rx SI vector $\boldsymbol{r}_{\circ}\in\mathbb{C}^{N_{\circ}}$ ($N_\circ=L_\circ-L+1$) can be expressed as
\begin{align}\label{vec_r}
    \boldsymbol{r}_\circ = \boldsymbol{\Phi}_\circ{\boldsymbol{w}}+\boldsymbol{\varepsilon}_\circ+\boldsymbol{z}_\circ,\ \circ\in\{\rm{p},\rm{d}\},
\end{align}
where the measurement matrix $\boldsymbol{\Phi}_\circ\in\mathbb{C}^{N_\circ\times L_w}$ is defined as
\begin{align}\label{Phi}
    \boldsymbol{\Phi}_\circ^\mathrm{T} = [\boldsymbol{\phi}\big(x_\circ[n]\big)\ \cdots\ \boldsymbol{\phi}\big(x_\circ[n+N_\circ-1]\big)].
\end{align}
Then, when the effective pilot length $N_{\rm p}\ge L_w$, the LS estimation of the weights can be expressed as
\begin{align}\label{w_ph_ls}
    \hat{\boldsymbol{w}}&=(\boldsymbol{\Phi}_\mathrm{p}^\mathrm{H}\boldsymbol{\Phi}_\mathrm{p})^{-1}\boldsymbol{\Phi}_\mathrm{p}^\mathrm{H} \boldsymbol{r}_\mathrm{p}\notag\\
    &=\boldsymbol{\Phi}_\mathrm{p}^{\dagger}(\boldsymbol{\Phi}_\mathrm{p}{\boldsymbol{w}}+\boldsymbol{\varepsilon}_\mathrm{p}+\boldsymbol{z}_\mathrm{p})\notag\\
    &=\boldsymbol{w}+\boldsymbol{\Phi}_\mathrm{p}^{\dagger}(\boldsymbol{\varepsilon}_\mathrm{p}+\boldsymbol{z}_\mathrm{p}).
\end{align}
Therefore, the RSI vector $\boldsymbol{e}_{\rm d}\in\mathbb{C}^{N_{\rm d}}$ for data sequences can be expressed as
\begin{align}\label{vec_e}
    \boldsymbol{e}_\mathrm{d}&=\boldsymbol{r}_\mathrm{d}-\boldsymbol{\Phi}_{\ds}\hat{\boldsymbol{w}}\notag\\
    &=\boldsymbol{\Phi}_{\ds}{\boldsymbol{w}}+\boldsymbol{\varepsilon}_{\ds}+\boldsymbol{z}_{\ds}-\boldsymbol{\Phi}_{\ds}\left(
    \boldsymbol{w}+\boldsymbol{\Phi}_\mathrm{p}^{\dagger}(\boldsymbol{\varepsilon}_\mathrm{p}+\boldsymbol{z}_\mathrm{p})
    \right)\notag\\
    &=\underbrace{\boldsymbol{\varepsilon}_\mathrm{d}}_{\text{Truncation error}} 
    -\underbrace{\boldsymbol{\Phi}_\mathrm{d}\boldsymbol{\Phi}_\mathrm{p}^{\dagger}\boldsymbol{\varepsilon}_\mathrm{p}}_{\substack{\text{Bias-induced } \\ {\text{reconstruction error}}} }
    -
    \underbrace{\boldsymbol{\Phi}_\mathrm{d}\boldsymbol{\Phi}_\mathrm{p}^{\dagger}\boldsymbol{z}_\mathrm{p}}_{\substack{\text{Noise-induced } \\ {\text{reconstruction error}}} }
    +\boldsymbol{z}_\mathrm{d},
\end{align}
which corresponds to: \textit{truncation error}, \textit{bias-induced reconstruction error (BIRE)}, \textit{noise-induced reconstruction error (NIRE)} and \textit{Rx noise}, respectively. Here, “\textit{bias-induced}” refers to the deterministic component of the reconstruction error induced by the truncated nonlinear components during channel training, which could become dominant when optimizing the noise-induced reconstruction error as analyzed later.

To facilitate the analysis of the reconstruction error, we resort to the GLP representation of PH cancellers, as described in Section \ref{GLP}.
\subsubsection{GLP expansion}
\ \newline
\indent The Rx SI vector in Eq. \eqref{vec_r} can be rewritten by the GLP representation, which is
\begin{align}
    \boldsymbol{r}_\circ = &\boldsymbol{\Psi}_\circ{\boldsymbol{\theta}}+\boldsymbol{\epsilon}_\circ+\boldsymbol{z}_\circ.
\end{align}
It can be proved that the PH canceller is equivalent to a GLP-based canceller under LS estimation. Specifically, we have: 
\begin{thm}\label{thm_1}
     \textup{(Equivalence of GLP \& PH Canceller)} The GLP canceller is equivalent to the PH canceller if LS estimation for the weights is employed for both cancellers, in the sense that
    \begin{align}
        \boldsymbol{\Phi}_\mathrm{d}\hat{{\boldsymbol{w}}} = &\boldsymbol{\Psi}_\mathrm{d}\hat{\boldsymbol{\theta}},
    \end{align}
    where
    \begin{align}
        \hat{{\boldsymbol{w}}}=&\boldsymbol{\Phi}_\mathrm{p}^\dagger\boldsymbol{r}_\mathrm{p}\notag\\
        \hat{\boldsymbol{\theta}}=&\boldsymbol{\Psi}_\mathrm{p}^\dagger\boldsymbol{r}_\mathrm{p}.
    \end{align}
    \begin{proof}
        See Appendix A.
    \end{proof}
\end{thm}
Therefore, the RSI vector in Eq. \eqref{vec_e} can be rewritten as
\begin{align}
    \boldsymbol{e}_\mathrm{d}=\boldsymbol{\epsilon}_\mathrm{d} -
    \boldsymbol{\Psi}_\mathrm{d}\boldsymbol{\Psi}_\mathrm{p}^{\dagger}\boldsymbol{z}_\mathrm{p}
    -\boldsymbol{\Psi}_\mathrm{d}\boldsymbol{\Psi}_\mathrm{p}^{\dagger}\boldsymbol{\epsilon}_\mathrm{p}
    +\boldsymbol{z}_\mathrm{d}.
\end{align}
Assuming that $x_{\rm d}[n]\overset{i.i.d.}{\sim}\mathcal{CN}(0,1)$, $\boldsymbol{z}_\mathrm{p}\sim\mathcal{CN}(\boldsymbol{0},\sigma_z^2\boldsymbol{I}_{N_{\ps}})$, and $\boldsymbol{z}_\mathrm{d}\sim\mathcal{CN}(\boldsymbol{0},\sigma_z^2\boldsymbol{I}_{N_{\ds}})$ are mutually independent, we have $\E{\boldsymbol{\Psi}^{\Ht}_{\ds}\boldsymbol{\epsilon}_{\ds}}=\boldsymbol{0}$ and $\E{\boldsymbol{\Psi}_{\ds}^{\Ht}\boldsymbol{\Psi}_{\ds}}=\E{\boldsymbol{G}_{\ds}}=N_{\ds}\boldsymbol{I}_{L_w}$, where the Gram matrix $\boldsymbol{G}_\circ = \boldsymbol{\Psi}_\circ^\mathrm{H}\boldsymbol{\Psi}_\circ$. Furthermore, the expected power of the RSI can be expressed as
\begin{align}\label{E_rho_RSI}
    \rho_e&=\mathbb{E}\bigg\{
    \frac{
    \norm{\boldsymbol{e}_{\ds}}^2}{N_{\ds}}\bigg\}\notag\\
    &=\frac{\E{\|\boldsymbol{\epsilon}_\mathrm{d}\|^2}}{N_{\rm d}}+ \|\boldsymbol{\Psi}_\mathrm{p}^\dagger\boldsymbol{\epsilon}_\mathrm{p}\|^2 
    + 
    \mathrm{tr}(\boldsymbol{G}_\mathrm{p}^{-1})\sigma_z^2+ \sigma_z^2,
\end{align}  
which corresponds to the power of the \textit{truncation error}, \textit{BIRE}, \textit{NIRE}, and \textit{Rx noise}, respectively. 

\subsection{Optimization criterion for the pilot sequence}
To minimize the RSI in Eq. \eqref{E_rho_RSI}, it suffices to minimize the sum of the NIRE and the BIRE powers, i.e.,
\begin{align}\label{opt_obj}
    \min_{x_{\ps}}\  J(\boldsymbol{x_{\ps}})=\norm{\boldsymbol{\Psi}_\mathrm{p}^\dagger\boldsymbol{\epsilon}_\mathrm{p}}^2 
    + 
    \mathrm{tr}(\boldsymbol{G}_\mathrm{p}^{-1})\sigma_z^2.
\end{align}
It is difficult to explicitly construct the pilot sequence based on $J(\boldsymbol{x}_{\ps})$ since the structure of the measurement matrix $\boldsymbol{\Psi}_{\ps}$ with nonlinear basis functions is quite complicated, and moreover we have no knowledge about the weights in the truncation error. To circumvent this difficulty, we resort to generating a candidate set of pilot sequences and then selecting the optimal one from this set based on a certain criterion. Therefore, the key issues here are: 1) How should we generate the candidate set? 2) Which criterion should be employed for choosing the optimal pilot sequence? 

To address the first issue, i.e., which candidate set could potentially achieve low RSI, we first explore the asymptotic case (infinite pilot length) and then move to the non-asymptotic case.
\subsubsection{The asymptotic optimality of Gaussian sequences}
\ \newline
\indent
In the asymptotic case, i.e., when the pilot  is of infinite length, it can be proven that the LS estimation is \textit{asymptotically unbiased} if $x_\mathrm{p}[n]\overset{i.i.d.}{\sim}\mathcal{CN}(0,1)$, which is stated as follows:
\begin{thm}\label{thm_2}
    \textup{(Asymptotic Unbiasedness by Gaussian Sequence)} If the pilot sequence is independently Gaussian distributed, the LS estimator of the channel coefficients $\boldsymbol{\theta}$ would be asymptotically \textit{unbiased}, namely,
    \begin{equation}
         \mathbb{E}_z\big\{\hat{\boldsymbol{\theta}}-\boldsymbol{\theta}|\ \boldsymbol{x}_\mathrm{p}\big\}=\boldsymbol{\Psi}_{\ps}^\dagger\epsilon_{\ps}\xrightarrow[N_\mathrm{p}\rightarrow\infty]{a.s.}\boldsymbol{0}.
    \end{equation}    
    \begin{proof}
        See Appendix B.
    \end{proof}
\end{thm}
\textbf{Theorem 2} implies that, if $x_\mathrm{p}[n]\overset{i.i.d.}{\sim}\mathcal{CN}(0,1)$, the expected power of the BIRE will asymptotically equal zero. Furthermore, the NIRE will also asymptotically equal zero since
\begin{align}
    \frac{1}{N_{\ps}}\boldsymbol{G}_{\ps}&=\frac{1}{N_{\ps}}\sum_{n}^{N_{\ps}}\boldsymbol{\psi}^*\big(x_{\ps}[n]\big)\boldsymbol{\psi}^{\rm  T}\big(x_{\ps}[n]\big)\notag\\
    &\xrightarrow[N_\mathrm{p}\rightarrow\infty]{a.s.}\E{\boldsymbol{\psi}^*\big(x_{\ps}[n]\big)\boldsymbol{\psi}^{\rm  T}\big(x_{\ps}[n]\big)}=\boldsymbol{I}_{L_w}.
\end{align}
This means that the Gaussian sequence is optimal in the asymptotic sense when the pilot sequence is of infinite length, resulting in $J(\boldsymbol{x}_{\ps})\rightarrow0$ as $N_\mathrm{p}\rightarrow\infty$. Then, a natural question is whether the Gaussian sequence is still optimal in the case of a finite pilot length.
\subsubsection{Trade-off between NIRE and BIRE}
\ \newline
\indent When the pilot is of finite length,  since both the NIRE and BIRE will be nonzero, we need to analyze them in more detail. 

For the NIRE part,  its key term $\tr{\boldsymbol{G}_{\ps}^{-1}}$  can be expressed as
\begin{align} \label{NIRE}
    \tr{\boldsymbol{G}_{\ps}^{-1}}&=\sum_{i=1}^{L_w}\frac{1}{\lambda_i(\boldsymbol{G}_{\ps})}\notag\\
    &=\frac{1}{\tr{\boldsymbol{G}_{\ps}}}\sum_{i=1}^{L_w}\frac{1}{\bar\lambda_i},
\end{align}
where $\bar\lambda_i=\lambda_i(\boldsymbol{G}_{\ps})/\tr{\boldsymbol{G}_{\ps}}$ is the normalized eigenvalue satisfying $\sum1/\bar\lambda_i\ge L_w^2$ for $\sum1/\bar\lambda_i= L_w^2$ when $\bar\lambda_1=\bar\lambda_2=\dots=\bar\lambda_{L_w}=1/L_w$. Meanwhile, since
\begin{align}
    \tr{\boldsymbol{G}_{\ps}}&=\tr{\sum_{n}^{N_{\ps}}\boldsymbol{\psi}^*\big(x_{\ps}[n]\big)\boldsymbol{\psi}^{\rm  T}\big(x_{\ps}[n]\big)}\notag\\
    &=\sum_{n}^{N_{\ps}}\Big\|
    \boldsymbol{\psi}\big(x_{\ps}[n]\big)
    \Big\|^2\notag\\
    &=\sum_{n}\sum_{q}\sum_{\ell}\Big|\psi_q\big(x_{\ps}[n-\ell]\big)\Big|^2=\Theta(u_{\max}^P)
\end{align}
is of the same order as $u_{\max}^P$, there may exist a pilot sequence with a higher PAPR than the Gaussian sequence, which is capable of obtaining a lower NIRE once the denominator in Eq. \ref{NIRE} $\tr{\boldsymbol{G}_{\ps}}$ grows faster than the numerator $\sum1/\bar\lambda_i$ (which grows with the increase of the degree of spectral non-uniformity).

However, it is not difficult to conclude that excessively high PAPR will lead to high BIRE, due to 1) the increase of the correlation between the truncation error $\boldsymbol{\epsilon}_{\ps}$ of PH and the measurement matrix $\boldsymbol{\Psi}$ which depends on the retained terms of PH; 2) the increase of the truncation error by at least $u_{\max}^{P+2}$, which grows faster than  $\tr{\boldsymbol{G}_{\ps}}$ since $\norm{\boldsymbol{\epsilon}_{\ps}}^2=\Omega(u_{\max}^{P+2})$. Therefore, to achieve the lowest RSI, the PAPR should be set to an appropriate value in order to strike a balance between the NIRE and the BIRE. Next, we analyze the effect of PAPR on NIRE in more detail, namely, demonstrating that a higher-PAPR pilot sequence can result in lower NIRE in a rigorous manner.

a) \textit{NIRE-oriented pilot design:}

The optimization criterion for NIRE-oriented pilot design is $\min_{\boldsymbol{x}_{\ps}} {\rm tr}(\boldsymbol{G}_{\rm p}^{-1})$, which corresponds to the  criterion proposed in \cite{Brink24} and is essentially based on the \textit{Cramér-Rao bound (CRB)}. 

As discussed above, high PAPR sequences may obtain a lower NIRE due to the strong excitation of high-order nonlinear basis functions by high PAPR. Thus, we consider the generalized Pareto distribution (GPD) with a heavier tail (implying higher PAPR) than the exponential distribution to generate the squared modulus of the pilot sequence, and the phase is generated by a uniform distribution, i.e., $x_{\ps}=\sqrt{u_n}e^{j\varphi_n}$ with 
$\varphi_n\overset{i.i.d.}{\sim}{\rm Unif}[0,2\pi)$ and $u_{n}\overset{i.i.d.}{\sim}{\rm GPD}(\xi,1-\xi)$, where $\E{u_n}=1$, and the cumulative distribution function (CDF) of ${\rm GPD}(\xi,1-\xi)$ can be expressed as
\begin{align}
    F_{\xi}(u)=1-\Big(1+\frac{\xi u}{1-\xi}\Big)^{-1/\xi}.
\end{align}
Particularly, when $\xi=0$, the GPD is equivalent to the exponential distribution, i.e., ${\rm GPD}(0,1)\sim{\rm Exp}(1)$. Compared to the exponential distribution, the GPD has a heavier tail when $\xi>0$, resulting in a higher probability of generating high-value samples, i.e., high PAPR sequences.

It can be proven that the sequence with GPD squared modulus could produce a lower asymptotic NIRE than the Gaussian sequence, which is stated as follows:
\begin{proposition}\label{proposition1}
    \textup{(NIRE Reduction via Heavier-Tailed Amplitude Distribution)} For an $i.i.d.$ random pilot sequence $x_{\rm p}(n)=\sqrt{u_{n}}e^{j\varphi_n}$, where $u_{n}\overset{i.i.d.}{\sim}{\rm GPD}(\xi,1-\xi)$ is independent of $\varphi_n\overset{i.i.d.}{\sim}{\rm Unif}[0,2\pi)$, there exists a random sequence which can achieve a smaller  ${\rm tr}(\boldsymbol{G}_{\rm p}^{-1})$ than the Gaussian sequence, which is expressed as

    \begin{align}
        \exists 0<\xi<\frac{1}{P}:\ {\rm tr}\big(\mathbb{E}\big\{\boldsymbol{G}_{\rm p}|\xi\big\}^{-1}\big)<{\rm tr}\big(\mathbb{E}\big\{\boldsymbol{G}_{\rm p}|0\big\}^{-1}\big).
    \end{align}  
    \begin{proof}
        See Appendix C.
    \end{proof}
\end{proposition}
\textbf{Proposition 1} indicates that, compared to the Gaussian sequences, the GPD sequences with a heavier-tailed amplitude for $\xi>0$ can obtain a lower asymptotic $\tr{\boldsymbol{G}_{\ps}^{-1}}$ since $\boldsymbol{G}_{\ps}\xrightarrow[N_\mathrm{p}\rightarrow\infty]{a.s.}\E{\boldsymbol{G}_{\ps}}$.

Furthermore, in the case of $L_{\ps}$-length pilot sequences with the power constraint $\sum \abs{x_{\rm p}(n)}^2=L_{\ps}$, to simplify the analysis, we first consider  $\boldsymbol{G_{\ps}}$ in the  memoryless scenario (i.e.,  $L=1$ and $L_w=(P+1)/2$), which can be expressed as a function of the empirical distribution of the squared modulus as follows:
\begin{align}
    \boldsymbol{G}_{\ps}(\boldsymbol{u})=\sum_{n}^{L_{\ps}}\boldsymbol{\psi}_P(\sqrt{u_n})\boldsymbol{\psi}_P^{\Ht}(\sqrt{u_n}),
\end{align}
where $\boldsymbol{u}$ is the vector of the squared modulus, and
\begin{align}
    \boldsymbol{\psi}_P(\sqrt{u_n})=\big[\psi_1(\sqrt{u_n})\ \cdots\ \psi_P(\sqrt{u_n}) \big]^{\rm T}.
\end{align}
While $\boldsymbol{G_{\ps}}$ is solely determined by the empirical distribution of the squared modulus $u_n$, we can construct a new sequence $x_{\rm p}(n,\delta)=\sqrt{u_{n,\delta}}e^{j\varphi_n}$ with higher PAPR by distributing a part of power of the elements in a given subset to the single element with the highest power, i.e., $u_{i,\delta}=u_i+\delta$ where $u_i=\max\limits_{n}u_n$, and $u_{j,\delta}=u_j-\frac{\delta}{M},\ j\in S_0$, where $M=|S_0|$ and $S_0=\{ j: \norm{\boldsymbol{\psi}_P(\sqrt{u_j})}^2\in \mathcal{U}_0, u_j>0 \}$. The sufficient condition for this new sequence to obtain a lower NIRE can be formulated as:
\begin{proposition}\label{proposition2}
    \textup{(NIRE Reduction via Increasing PAPR through Power Redistribution)} For the reconstructed pilot sequence $x_{\rm p}(n,\delta)$, we can find a power-perturbed  pilot sequence with higher PAPR and lower NIRE if the following condition is satisfied:
    \begin{equation}
        \gamma=\frac{d\tr{\boldsymbol{G}_{\ps}^{-1}(\boldsymbol{u}_{\delta})}}{d\delta}\Bigg|_{\delta=0}\notag <0\\
    \end{equation}
    \begin{proof}
        See Appendix D.
    \end{proof}
\end{proposition}
\begin{figure}[!t]
\centering
\includegraphics[width=3.2in]{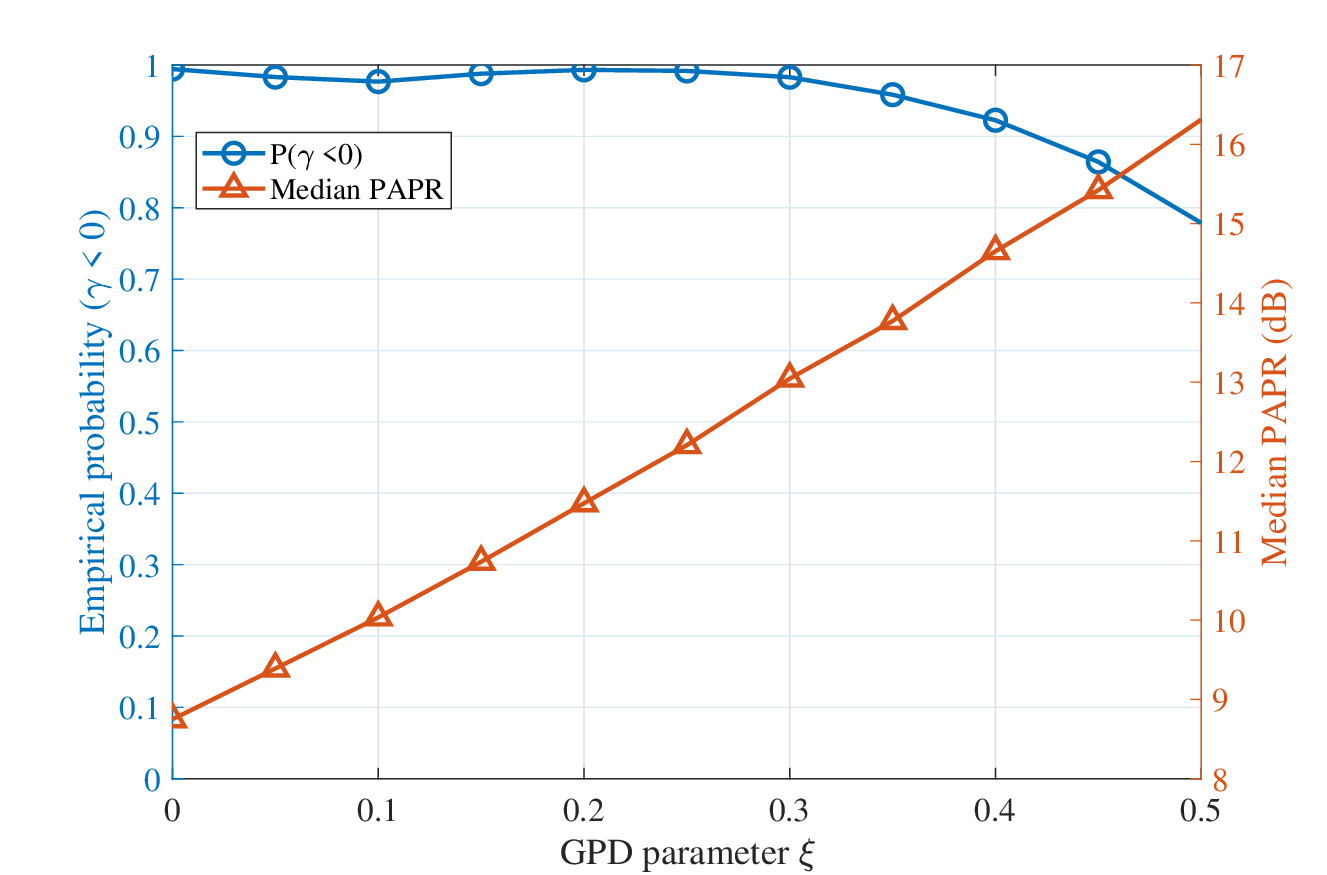}
\caption{The empirical probability of $\gamma<0$ and the median PAPR over 10000 GPD sequences with different shape parameters. ($L_\mathrm{p}=1280,P=7,L=1$)}
\label{Pr_xi_PAPR}
\end{figure}
Fig. \ref{Pr_xi_PAPR} demonstrates the empirical probability of $\gamma<0$ under 10000 GPD sequences with different parameters $\xi$, where $S_0=\{ j: \norm{\boldsymbol{\psi}_P(\sqrt{u_j})}^2\in[0.5,1], u_j>0 \}$. It can be seen that even though the PAPR is already very high, there is still a high probability of reducing $\tr{\boldsymbol{G}_{\ps}^{-1}}$ by increasing the PAPR of the pilot sequence since the energy of the GLP basis vector $\boldsymbol{\psi}_P(\sqrt{u})$ will increase asymptotically by $u^P$.

\begin{figure}[!t]
\centering
\includegraphics[width=3.0in]{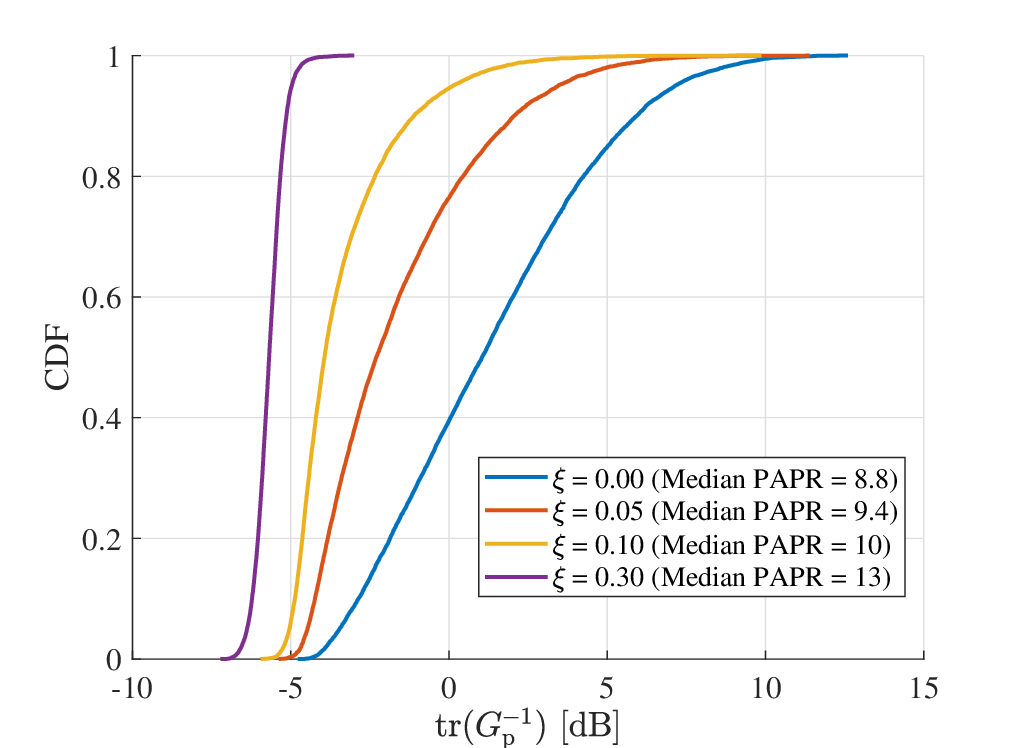}
\caption{The CDF of ${\rm tr}(\boldsymbol{G}_{\ps}^{-1})$ under GPD sequences at different $\xi$. ($L_\mathrm{p}=1280,P=7,L=100$.)}
\label{CDF_TrinvG}
\end{figure}

Compared to the memoryless case, the Gram matrix $\boldsymbol{G}_{\ps}$ in the memory case has an additional element of the cross-correlation of the GLP vectors with different delays, which can be expressed as
\begin{align}\label{cross-correlation}
    \sum_{n}^{N_p}\psi_q^*\big(x_{\ps}[n+m]\big)\psi_p\big(x_{\ps}[n+m-\ell]\big)
\end{align}
for $m=0,\dots,L-1$, and $\ell=1,\dots,L-1$. Although an exact characterization of the  $\boldsymbol{G}_{\ps}$ with memory is difficult due to the delay-induced cross-correlations, the orthogonality in Eq. \eqref{op_GLP} suggests that these cross-correlation terms tend to be small for independent and circularly symmetric pilot sequences. Accordingly, the preference of the \textit{CRB criterion} for the pilot sequence with high PAPR could also be valid for the memory $\boldsymbol{G}_{\ps}$. To validate this, Fig. \ref{CDF_TrinvG} demonstrates the empirical CDF of $\tr{\boldsymbol{G}_{\ps}^{-1}}$ under 10000 GPD sequences at different parameters $\xi$. It can be seen that as $\xi$ increases from $0$ to $0.3$, the median PAPR rises from $8.8$ dB to $13$ dB, while the entire CDF of $\tr{\boldsymbol{G}_{\ps}^{-1}}$ shifts monotonically to the left. This indicates that the reduction mechanism predicted by \textbf{Proposition 2} under memoryless $\boldsymbol{G}_{\ps}$ remains statistically effective in the case of memory $\boldsymbol{G}_{\ps}$.

The above analyses indicate that compared to Gaussian sequences, GPD sequences with higher PAPR due to the heavy-tailed distribution can obtain a lower NIRE, which reveals that high PAPR is desirable for pilot sequences in nonlinear estimation. However, it is not always beneficial to achieve a lower NIRE with a pilot sequence of higher PAPR, especially when the PH order $P$ is not sufficiently large.

b) \textit{BIRE penalty caused by high-PAPR pilot sequence}:

Regarding the BIRE, it can be bounded by the 2-norm inequality $\norm{\boldsymbol{A}\boldsymbol{x}}_2\le\norm{\boldsymbol{A}}_{\rm 2}\norm{\boldsymbol{x}}_2$ as follows:
\begin{align}\label{BIRE_upperbound1}
    \norm{\boldsymbol{\Psi}_{\ps}^\dagger\boldsymbol{\epsilon}_{\ps}}^2=\norm{\boldsymbol{G}_{\ps}^{-1}\boldsymbol{\Psi}_{\ps}^{\Ht}\boldsymbol{\epsilon}_{\ps}}^2
    &\le
    \norm{\boldsymbol{G}_{\ps}^{-1}}_2^2\norm{\boldsymbol{\Psi}_{\ps}^{\Ht}\boldsymbol{\epsilon}_\mathrm{p}}^2\notag\\
    &\le\norm{\boldsymbol{G}_{\ps}^{-1}}_2^2\norm{\boldsymbol{\Psi}_{\ps}^{\Ht}}_2^2\norm{\boldsymbol{\epsilon}_\mathrm{p}}^2\notag\\
    &=\frac{\lambda_{\max}(\boldsymbol{G}_\mathrm{p})}{\lambda_{\min}^2(\boldsymbol{G}_\mathrm{p})}\norm{\boldsymbol{\epsilon}_\mathrm{p}}^2\notag\\
    &=
    \frac{\kappa(\boldsymbol{G}_\mathrm{p})}{\lambda_{\min}(\boldsymbol{G}_\mathrm{p})}\norm{\boldsymbol{\epsilon}_\mathrm{p}}^2,
\end{align}
where $\kappa(\boldsymbol{G}_\mathrm{p})=\lambda_{\max}(\boldsymbol{G}_\mathrm{p})/\lambda_{\min}(\boldsymbol{G}_\mathrm{p})$, and the norm of the truncation error vector $\boldsymbol{\epsilon}_\mathrm{p}$ can be expressed as
\begin{align}
   \norm{\boldsymbol{\epsilon}_\mathrm{p}}^2&=\sum_{n}^{N_{\ps}}\left|\sum\limits_{\substack{q=P+2\\q={\rm odd}}}^{\infty}\sum\limits_{\ell=0}^{L-1}\theta_{q,\ell}\psi_q\big(x_\mathrm{p}[n-\ell]\big)\right|^2\notag\\
    &=\Omega\big(u_{\max}^{P+2}\big).
\end{align}
Particularly, if $x_\mathrm{p}[n]\overset{i.i.d.}{\sim}\mathcal{CN}(0,1)$, due to the orthogonality of GLP, we have
\begin{align}
    \E{\boldsymbol{\Psi}_{\ps}^{\Ht}\boldsymbol{\epsilon}_\mathrm{p}}=\boldsymbol{0},
\end{align}
while this does not hold for GPD since
\begin{align}
    \E{\psi_q^*(x)\psi_p(x)}
    =&\sum_{\substack{i=1\\i={\rm odd}}}^q
    \sum_{\substack{j=1\\j={\rm odd}}}^p
    \ell^*_{q,i}\ell_{p,j}\E{u_n^{(i+j)/2}}\notag\\
    =&\sum_{i}^q
    \sum_{j}^p
    \ell^*_{q,i}\ell_{p,j}\frac{\left(\frac{i+j}{2}\right)!(1-\xi)^{\frac{i+j}{2}}}{\prod_{k=1}^{\frac{i+j}{2}}(1-k\xi)},
\end{align}
is generally non-zero when $\xi\neq 0$. This means that the empirical distribution of the squared modulus with an excessively high PAPR could not only increase $\kappa(\boldsymbol{G}_{\ps})$ as well as the correlation between the unmodeled truncated subspace and the estimation subspace $\boldsymbol{\Psi}_{\ps}$, but also strengthen the energy of the truncation error, which increases by at least $u^{P+2}_{\max}$ compared with $\tr{\boldsymbol{G}_{\ps}}=\Theta(u^{P}_{\max})$. This means that if the PAPR increases by $1$ dB, the truncation error for $P=7$ could increase by $9$ dB.

Based on the above analysis, roughly speaking, compared to the Gaussian benchmark, the decrease in the NIRE by increasing the PAPR is at the cost of the increase in the BIRE, and there exists a trade-off strongly correlated with PAPR between the NIRE and the BIRE.

\subsubsection{Optimization criterion}
\ \newline
\indent Accordingly, the NIRE-oriented pilot design---\textit{CRB criterion} could perform well in the case of PH cancellers with a sufficiently large order since the BIRE is much weaker than the NIRE even if the pilot PAPR is extremely high. However, in practical systems, the PH order should be as small as possible while satisfying performance requirements. In this case, the impact of an excessively high PAPR on the BIRE needs to be carefully weighed in the optimization criterion.

Since $\tr{\boldsymbol{G}_{\ps}^{-1}}$ can be bounded as
\begin{align}
    \tr{\boldsymbol{G}_{\ps}^{-1}}=\sum_{i=1}^{L_w}\frac{1}{\lambda_i(\boldsymbol{G}_{\ps})}&\le\frac{L_w}{\lambda_{\min}(\boldsymbol{G}_{\ps})}\notag\\
    &\le\frac{L_w\kappa(\boldsymbol{G}_{\ps})}{\lambda_{\min}(\boldsymbol{G}_{\ps})},
\end{align}
according to the upper bound of the BIRE in Eq. \eqref{BIRE_upperbound1}, the optimization criterion in Eq. \eqref{opt_obj} can be bounded as:
\begin{align}
    \min_{x_{\ps}}\  J(\boldsymbol{x_{\ps}})&=\norm{\boldsymbol{\Psi}_\mathrm{p}^\dagger\boldsymbol{\epsilon}_\mathrm{p}}^2 
    + 
    \mathrm{tr}(\boldsymbol{G}_\mathrm{p}^{-1})\sigma_z^2\notag\\
    &\le\frac{\kappa(\boldsymbol{G}_{\ps})}{\lambda_{\min}(\boldsymbol{G}_{\ps})}(\norm{\boldsymbol{\epsilon}_\mathrm{p}}^2+L_w\sigma_z^2).
\end{align}
Therefore, a robust criterion taking into account both the NIRE and the BIRE, i.e., condition-number-to-minimum-eigenvalue ratio (\textit{CMER}), can be expressed as:
\begin{align}
    \min_{\boldsymbol{x}_{\ps}\in\mathcal{X}_{\ps}}\ &\frac{\kappa(\boldsymbol{G}_{\ps})}{\lambda_{\min}(\boldsymbol{G}_{\ps})}\\
    {\rm s.t.}\ \  &\norm{\boldsymbol{x}_{\ps}}^2=L_{\ps}.\notag 
\end{align}
Here, the candidate set $\mathcal{X}_{\ps}$ is defined as:
\begin{align}
    \mathcal{X}_{\ps}&=\bigcup_{\xi\in\Xi}\mathcal{X}_{\xi},
\end{align}
where $\Xi$ is the set of shape parameters $\xi$ and
\begin{align}
    \mathcal{X}_{\xi}=\Bigg\{&
    \boldsymbol{x}_{\ps}\in\mathbb{C}^{L_{\ps}}\Bigg|\ x_{\ps}[n]=\frac{\sqrt{L_{\ps}}x[n]}{\sqrt{\sum \abs{x[n]}^2}},x[n]=\sqrt{u_n}e^{j\varphi_n},\notag\\ &u_n\overset{i.i.d.}{\sim}{\rm GPD}(\xi,1-\xi),\varphi_n\overset{i.i.d.}{\sim}{\rm Unif}[0,2\pi)
    \Bigg\}.
\end{align}
It is worth noting that the \textit{CMER criterion} poses an implicit constraint on the PAPR, which is
\begin{align}\label{soft_PAPR_constraint}
    \frac{\kappa(\boldsymbol{G}_{\ps})}{\lambda_{\min}(\boldsymbol{G}_{\ps})}=\frac{\lambda_{\max}(\boldsymbol{G}_\mathrm{p})}{\lambda_{\min}^2(\boldsymbol{G}_\mathrm{p})}\ge&
    \frac{\max_{i} [\boldsymbol{G}_{\rm p}]_{ii}}{\big(\min_i [\boldsymbol{G}_{\rm p}]_{ii}\big)^2}\notag\\
    \ge&
    \frac{\abs{\psi_P(\sqrt{u_{\max}})}^2}{(\sum|x_{\ps}[n]|^2)^2}\notag\\
    =&\frac{\Theta(u_{\max}^P)}{L_{\ps}^2}.
\end{align}

\subsection{Key Factors for PH Canceller}
According to Eq. \eqref{E_rho_RSI}, the performance of the LS-based PH canceller is jointly determined by the truncation error, the reconstruction error, and the equivalent Rx noise. The equivalent Rx noise level depends on the A-SIC capability, Tx noise, Rx noise, and ADC quantization noise. Therefore, A-SIC should first reduce the SI to within the ADC dynamic range and the linear region of the LNA; otherwise, D-SIC will be limited by front-end distortion and noise rather than by the digital cancellation model itself. Specifically, practical D-SIC design should consider multiple factors as follows:
\begin{itemize}
    \item \textit{A-SIC}: On the one hand, A-SIC needs to reduce SI to a sufficiently low level to prevent truncation distortion due to ADC saturation and nonlinear distortion generated by the LNA that cannot be modeled by the PH canceller. On the other hand, A-SIC determines the strength of Tx noise in the residual SI entering D-SIC, and the strength of equivalent quantization noise is controlled by the gain of the Rx chain determined by the dominant term between the residual SI before D-SIC and the expected signal.
    \item \textit{Order of the PH canceller}: Increasing the PH order reduces the truncation error but also increases the number of weights, which could enlarge the reconstruction error with a certain pilot. Therefore, it can be inferred that there exists an optimal order to achieve the optimal balance between the truncation error and the reconstruction error.
    \item \textit{The power delay profiles of SI channels and the bandwidth of SI}: The power delay profiles of SI channels and the bandwidth of SI jointly determine the memory length $L$ of PH cancellers. A larger channel delay spread or signal bandwidth requires a longer memory length and hence a longer pilot length. Otherwise, the reconstruction error could increase as the number $L_w$ of weights increases.
    \item \textit{Multiple antennas}: On the one hand, multiple antennas mean that the pilot length increases with the number of antennas; On the other hand, the input power of the PA in each Tx chain will decrease under restricted total Tx power, resulting in weaker PA nonlinearity and a decrease in the required PH order. This means that multiple antennas and PA efficiency need to be considered jointly to achieve a balance in practical systems.
\end{itemize}

\section{Simulations and Discussion}
In this section, we validate the proposed pilot-design criterion through simulations based on an 80 MHz 802.11ax IBFD setting. The simulations compare different pilot sequences and examine how the selection criterion, candidate set, PAPR constraint, and pilot length affect the BIRE--NIRE trade-off and the final RSI.

\subsection{Simulation Settings}
The main simulation parameters are summarized in Table \ref{IBFD radio}. We consider an 80 MHz 802.11ax high-efficiency single-user (HE-SU) IBFD system and use the Rayleigh SI channel with $100$ taps to model the SI channel. The PA nonlinearity is modeled by the Rapp AM-AM characteristic and the AM-PM characteristic in Eqs. \eqref{AM-AM} and \eqref{AM-PM} as follows, with the parameters listed in Table \ref{NL model}.
\begin{align}\label{AM-AM}
    \mathrm{R}(x)=\frac{Ax}{\Big(1+\Big|\frac{Ax}{A_{\mathrm{sat}}}\Big|^{2s}\Big)^{\frac{1}{2s}}},
\end{align}
\begin{align}\label{AM-PM}
    \Theta(x)=\exp\Bigg(j\frac{B|x|^q}{1+\Big|\frac{x}{B_{\mathrm{sat}}}\Big|^{q}}\Bigg).
\end{align}
Here, $A$ is the linear gain of the PA, $A_{\mathrm{sat}}$ is the amplitude saturation level, $s$ is the amplitude smoothing factor, $B$ is the phase gain, $B_{\mathrm{sat}}$ is the phase saturation level, and $q$ is the phase smoothing factor.
\begin{table}[!t]
\caption{Configurations\label{IBFD radio}}
\centering
\begin{tabular}{c|c}
\hline
Parameter & Value\\
\hline
Standard &  802.11 ax (HE SU)\\
\hline
Bandwidth & 80 MHz\\
\hline
Modulation and coding scheme (MCS) & 64 QAM 3/4\\
\hline
Tx power $\rho_{\tx}$ & 23 dBm\\
\hline
Antenna number & 1\\
\hline
SI channel model & Rayleigh\\
\hline
A-SIC performance & 60 dB\\
\hline
Memory length $L$ & 100\\
\hline
Rx noise power $\sigma_z^2$ & -90 dBm\\
\hline
Trials & 1000\\
\hline
\end{tabular}
\end{table}

\begin{table}[!t]
\caption{Parameters of the nonlinear model \label{NL model}}
\centering
\begin{tabular}{c|c}
\hline
Parameter & Value\\
\hline
Linear gain $|A|^2$ &  30 dB\\
\hline
Amplitude saturation power $|A_{\mathrm{sat}}|^2$ & 30 dBm\\
\hline
Amplitude smoothing factor $s$ & 2\\
\hline
Phase gain $B$ & -0.15\\
\hline
Phase saturation level $B_{\mathrm{sat}}$ & 0.88\\
\hline
Phase smoothing factor $q$ & 2\\
\hline
\end{tabular}
\end{table}

\subsection{Performance Validation}
First, the optimized binary phase-shift keying (BPSK) orthogonal frequency-division multiplexing (OFDM) is designed by the Algorithm 1 in \cite{Brink24}, and the optimized GPD sequence is selected from $20000$ candidate GPD sequences with normalized power for $\xi=0,\ 0.05,\ 0.1,\ 0.3$, i.e., $\mathcal{X}_{\ps}^{(0)}=\bigcup_{\xi\in\Xi}\mathcal{X}_{\xi}$, where $\Xi=\{0,0.05,0.1,0.3\}$ and $|\mathcal{X}_{\xi}|=5000$. In addition, the ideal benchmark is the least-squares of the RSI obtained by estimating the weight with the data sequence as the pilot sequence, i.e., $\boldsymbol{x}_\mathrm{p}=\boldsymbol{x}_\mathrm{d}$.
\subsubsection{Superiority of the optimized GPD sequence}
\begin{figure}[!t]
\centering
\includegraphics[width=2.8in]{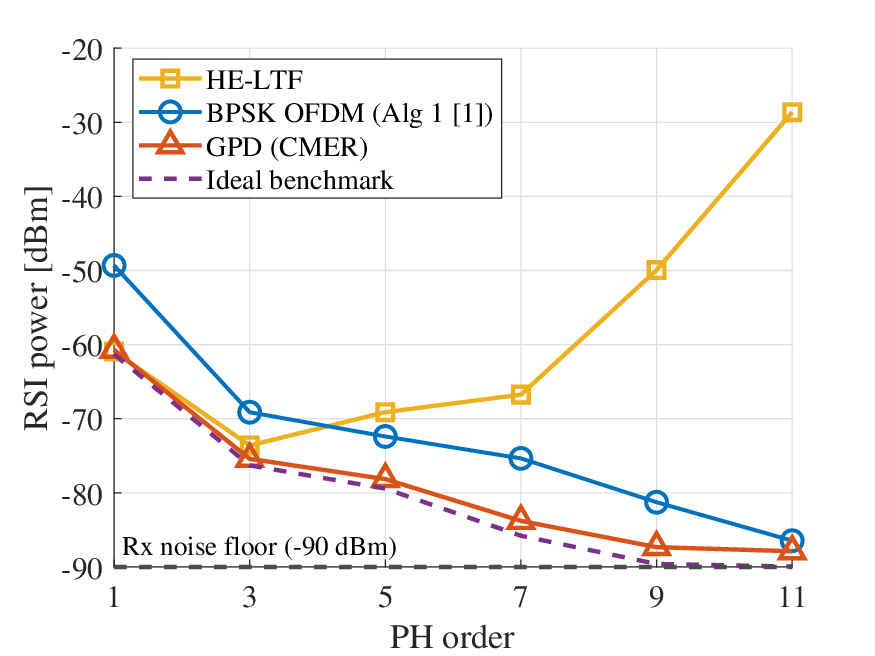}
\caption{The power of the RSI versus PH orders under several pilot sequences. ($L_\mathrm{p}=1280$)}
\label{RSIvsP_Seq}
\end{figure}
Fig. \ref{RSIvsP_Seq} illustrates the RSI power based on the PH canceller with different orders under the conventional HE-LTF, the optimized BPSK-OFDM, and the optimized GPD sequence, where the length of all pilot sequences is one OFDM symbol, i.e., $L_\mathrm{p}=1280$ for $80$ MHz 802.11ax HE-SU physical-layer protocol data unit.

It can be observed that the optimized GPD pilot sequence achieves significantly lower RSI than HE-LTF and optimized BPSK-OFDM when $P\le11$, and is close to the ideal benchmark. Particularly, when $P=9$, the RSI with the optimized GPD pilot is about $-87.3$ dBm, close to the ideal benchmark of $-89.6$ dBm, while the RSI with optimized BPSK-OFDM is merely $-81.2$ dBm, which is $6.1$ dB higher than that with the optimized GPD pilot. 

Next, we will reveal the key factor behind the optimization of the pilot sequence---PAPR, from three aspects: a) Optimization criteria; b) Candidate sets; c) PAPR constraints. Considering that the ideal benchmark of RSI has already reached $-89.6$ dBm when $P=9$, comparable to the Rx noise floor with $-90$ dBm, further increasing the PH order has a marginal diminishing effect on reducing the RSI. Therefore, from the perspective of implementation complexity, we will investigate the PH canceller with $P=9$ in the following simulations.

\subsubsection{\textit{CMER criterion} vs. \textit{CRB criterion}}
\begin{figure}[!t]
\centering
\includegraphics[width=2.8in]{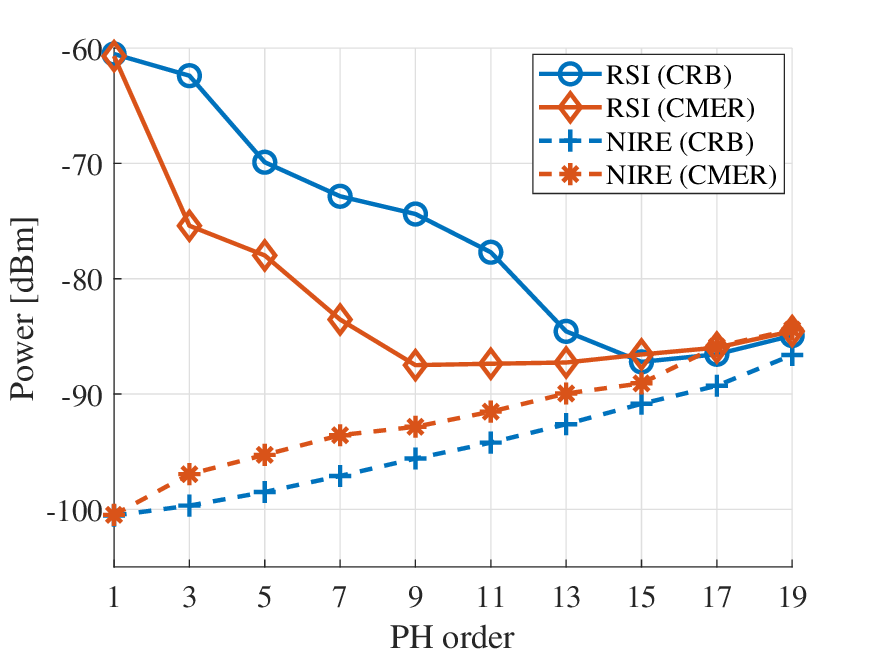}
\caption{The power of the RSI and the NIRE versus PH orders under the optimized GPD pilot sequence based on \textit{CMER} and \textit{CRB criteria}. ($L_\mathrm{p}=1280$)}
\label{RSI_NIREvsP_Criteria}
\end{figure}
To fairly compare the \textit{CRB} and \textit{CMER criteria}, the corresponding pilot sequences are selected from the same candidate set $\mathcal{X}_{\ps}^{(0)}$, and Fig. \ref{RSI_NIREvsP_Criteria} illustrates the power of the RSI and the NIRE versus different PH orders under the different optimized GPD sequences based on \textit{CRB} and \textit{CMER criteria}. It can be seen that when $P\le13$, the RSI under GPD pilot based on the \textit{CMER criterion} is still lower than that based on the \textit{CRB criterion}, where the gap can even be as high as 13 dB at $P=9$, even though the GPD pilot based on the \textit{CRB criterion} can obtain a lower NIRE. As analyzed in Section \uppercase\expandafter{\romannumeral3}, the \textit{CRB criterion} has a preference for high PAPR sequences to obtain a lower NIRE. As a result, when the PH order is not sufficiently large, an excessively high PAPR will lead to a rapid increase in the BIRE, which becomes the dominant term in the RSI.

By contrast, the proposed \textit{CMER criterion} jointly considers the NIRE and the BIRE, and implicitly incorporates the PAPR constraint into the criterion as expressed in Eq. \eqref{soft_PAPR_constraint}. Although the pilot sequence based on the \textit{CRB criterion} can obtain a lower RSI when $P\ge15$, the gap of the RSI under different pilots based on the two criteria is no more than $1$ dB. 

\subsubsection{GPD candidate sets with different shape parameters}
\begin{figure}[!t]
\centering
\includegraphics[width=2.8in]{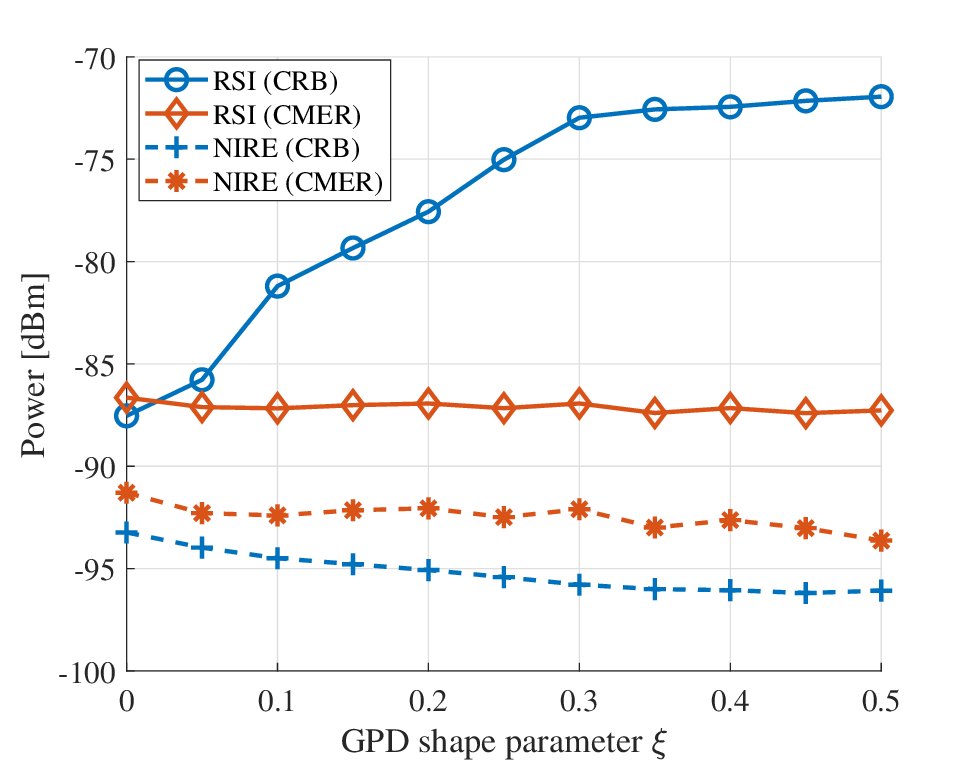}
\caption{The power of the RSI and the NIRE under the optimized GPD pilot sequence selected from the GPD candidate set with different shape parameters $\xi$ based on \textit{CMER} and \textit{CRB criteria}. ($P=9,\ L_{\ps}=1280$)}
\label{RSI_NIREvsPara_Criteria}
\end{figure}
To further evaluate the impact of candidate sets, Fig. \ref{RSI_NIREvsPara_Criteria} illustrates the power of the RSI and the NIRE with an 9-order PH canceller under the pilot sequences selected from GPD candidate sets with different parameters $\xi$, i.e., $\boldsymbol{x}_{\ps}\in\mathcal{X}_{\xi}$ for $0\le \xi \le 0.5$ and $|\mathcal{X}_{\xi}|=20000$. Since the probability of higher PAPR sequences generated in the candidate set increases as $\xi$ increases, the \textit{CRB criterion} without a hard PAPR constraint will lead to an increasingly severe imbalance between the NIRE and the BIRE due to its preference for high PAPR sequences. By contrast, the \textit{CMER criterion} is more robust to the change in the candidate set (PAPR) due to its soft PAPR constraint.  

It is worth noting that when $\xi=0$, i.e., a Gaussian candidate set, the pilot sequence based on the \textit{CRB criterion} can obtain a lower RSI than that based on the \textit{CMER criterion} since it achieves a better trade-off between the BIRE and the NIRE with the PAPR constraint inherent in the Gaussian candidate set. 

\subsubsection{Hard PAPR constraints}
\begin{figure}[!t]
\centering
\includegraphics[width=2.8in]{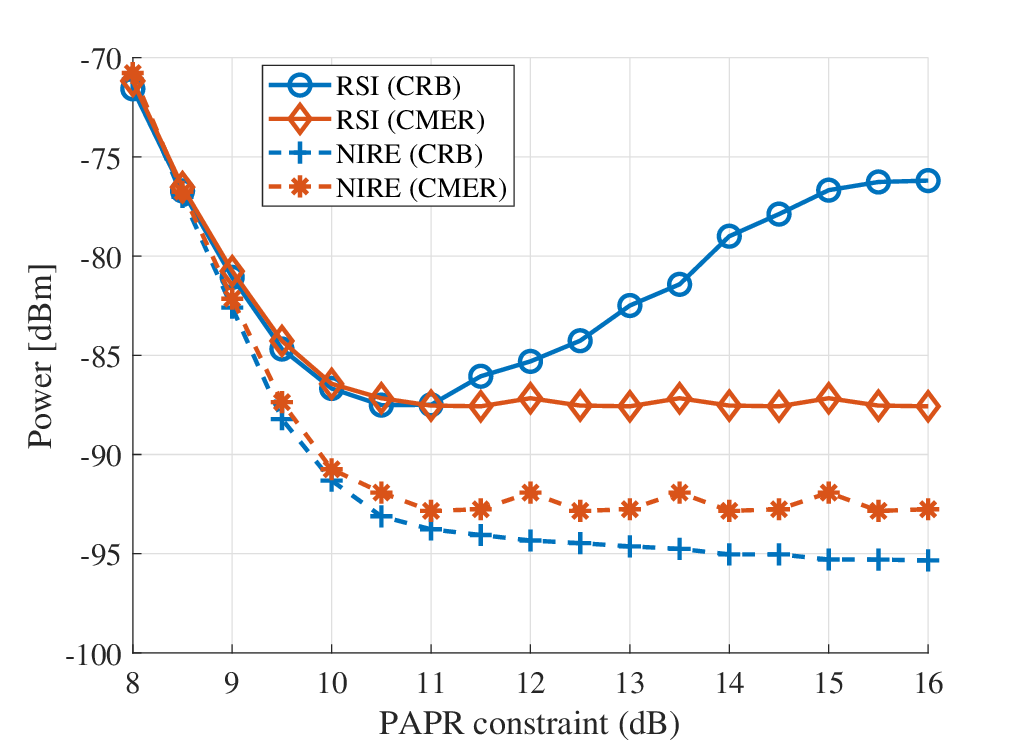}
\caption{The power of the RSI and the NIRE under the optimized GPD pilot sequence based on \textit{CMER} and \textit{CRB criteria} with different PAPR constraints. ($P=9,\ L_\mathrm{p}=1280$)}
\label{RSI_NIREvsPAPR_Criteria}
\end{figure}
To further demonstrate the impact of PAPR constraints, Fig. \ref{RSI_NIREvsPAPR_Criteria} illustrates the power of the RSI and the NIRE under GPD pilot sequences selected from $\mathcal{X}_{\ps}^{(0)}$ based on \textit{CMER} and \textit{CRB criteria} with hard PAPR constraints $\norm{\boldsymbol{x}_{\ps}}_\infty^2\le u_{\rm c}\in[8,16]$ dB. It can be clearly seen that when the PAPR constraint is no more than 11 dB, the RSI power decreases with the increase of the PAPR constraint for both pilot sequences based on \textit{CMER} and \textit{CRB criteria}, and the RSI power under the pilot based on \textit{CRB criterion} is even slightly lower than that based on \textit{CMER criterion}. By contrast, when the PAPR constraint is larger than 11 dB, the performance of the pilot based on the CRB criterion will sharply degrade as the PAPR constraint increases.

However, this hard PAPR constraint may not be suitable for other PA models and PH orders. Instead, the \textit{CMER criterion} with the soft PAPR constraint can adapt to various conditions while maintaining a rather good performance.

\subsubsection{BPSK OFDM vs. GPD}
\begin{figure}[!t]
\centering
\includegraphics[width=2.8in]{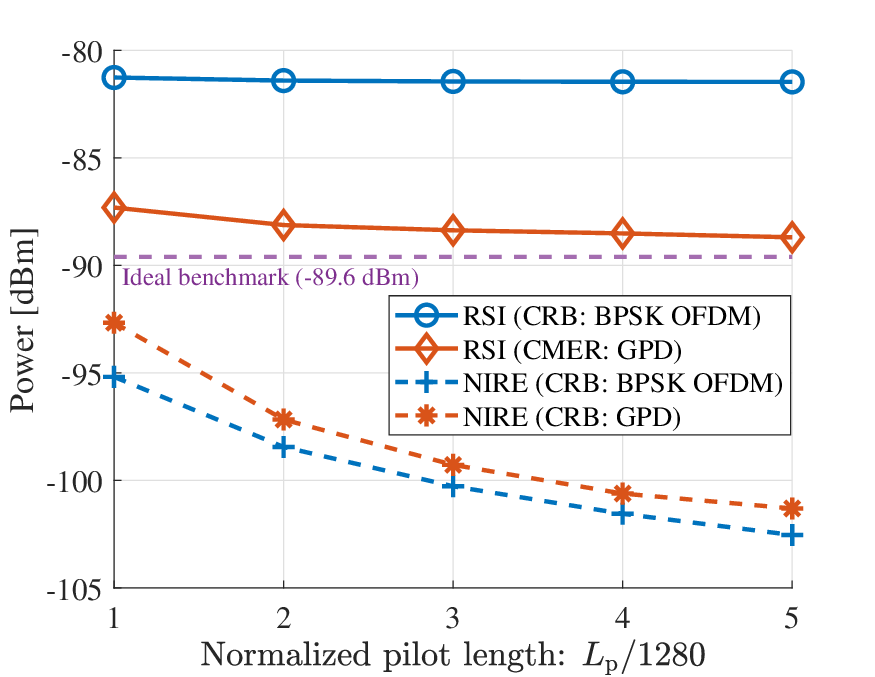}
\caption{The power of the RSI and the NIRE under the repeated BPSK OFDM (CRB) and the optimized GPD sequence (\textit{CMER}) with different pilot length. ($P=9$)}
\label{RSI_NIREvsLp_Seq}
\end{figure}
Although the BPSK OFDM pilot sequence is compatible with the 802.11ax standard, this structure also becomes a constraint on the performance of PH cancellers since the BIRE is a structural bias that does not decrease unless the correlation between the truncation error and the estimation subspace $\boldsymbol{\Psi}_{\ps}$ degrades.

Fig. \ref{RSI_NIREvsLp_Seq} illustrates the power of the RSI and the NIRE under the two pilot sequences when the pilot length $L_{\ps}$ increases, where the BPSK OFDM pilot with multiple OFDM symbols is obtained by repeating the single BPSK OFDM symbol and the GPD pilot sequence is selected from the candidate set with the corresponding pilot length. It can be seen that the RSI power under the GPD pilot sequence decreases from $-87.3$ dBm to $-88.7$ dBm as the pilot length increases from $1280$ to $5\times1280$, while the RSI power under BPSK OFDM is almost unchanged even though the NIRE decreases since the BIRE does not decrease with the number of repetitions $R$, which can be expressed as
\begin{align}
    (R\boldsymbol{G}_{\ps})^{-1}(R\boldsymbol{\Psi}_{\ps}^{\Ht}\boldsymbol{\epsilon}_{\ps})=\boldsymbol{G}_{\ps}^{-1}\boldsymbol{\Psi}_{\ps}^{\Ht}\boldsymbol{\epsilon}_{\ps},
\end{align}
where $\boldsymbol{G}_{\ps}$ and $\boldsymbol{\Psi}_{\ps}^{\Ht}\boldsymbol{\epsilon}_{\ps}$ only scale proportionally with $R$.

\subsection{Discussion}
The above results verify that the final RSI is not determined by the NIRE alone, but by the joint effect of truncation error, BIRE, NIRE, and Rx noise. Although the \textit{CRB criterion} can reduce the NIRE by favoring high-PAPR pilots, it may significantly increase the BIRE when the PH order is insufficient. This explains why a lower NIRE does not necessarily lead to a lower RSI.

By contrast, the proposed \textit{CMER criterion} provides a more balanced pilot selection. It preserves sufficient excitation of the nonlinear basis functions while avoiding excessive PAPR-induced BIRE. Therefore, \textit{CMER} achieves more robust RSI performance across different PH orders, GPD candidate sets, and PAPR constraints.

The comparison with repeated BPSK-OFDM pilots further shows that simply increasing the pilot length cannot effectively reduce the structural BIRE if the correlation between the truncated components and the estimation subspace remains unchanged. Hence, practical pilot design for nonlinear D-SIC should jointly consider the pilot distribution, PAPR, PH order, and training overhead.

\section{Conclusion}
This paper investigates the performance limit of the LS-based PH canceller for D-SIC in IBFD radios. Unlike conventional analyses that mainly focus on the NIRE, we incorporate the truncation-induced bias into the RSI analysis. By transforming the PH representation into an equivalent GLP representation, the RSI can be decomposed into the truncation error, the BIRE, the NIRE, and the Rx noise.

This framework reveals the fundamental BIRE--NIRE trade-off in pilot sequence design. With Gaussian pilots as an asymptotically optimal benchmark, high-PAPR pilots can further reduce the NIRE by strengthening the Gram matrix, but they may also increase the BIRE by enhancing the pilot truncation error and the correlation between the pilot truncation error and the estimation subspace. Therefore, the final RSI is not minimized by the NIRE-oriented \textit{CRB criterion} alone, especially when the PH order is limited.

Motivated by this trade-off, we propose the \textit{CMER criterion}, which balances the spectral strength and spectral uniformity of the Gram matrix and implicitly controls the PAPR of the selected pilot. Simulations based on an 80 MHz 802.11ax IBFD verify that the CMER-selected GPD pilot achieves lower RSI than HE-LTF and optimized BPSK-OFDM pilots, and provides more robust performance than the \textit{CRB criterion} in the BIRE-limited regime.

{\appendices
\section*{Appendix A\\Proof of \textbf{Theorem 1}}

According to Eq. \eqref{PH2GLP}, the measurement matrix of PH cancellers and its GLP representation satisfy the following relationship:
\begin{align}
    \boldsymbol{\Psi}_{\circ} = \boldsymbol{\Phi}_{\circ}\boldsymbol{T}.
\end{align}
Then, the LS estimation of the weights for the PH canceller can be expressed as
\begin{align}\label{W_W}
    \hat{\boldsymbol{w}}=&(\boldsymbol{\Phi}_\mathrm{p}^\mathrm{H}\boldsymbol{\Phi}_\mathrm{p})^{-1}\boldsymbol{\Phi}_\mathrm{p}^\mathrm{H}\boldsymbol{r}_\mathrm{p}\notag\\
    =&(\boldsymbol{T}^{-\Ht}\boldsymbol{\Psi}_\mathrm{p}^\mathrm{H}\boldsymbol{\Psi}_\mathrm{p}\boldsymbol{T}^{-1})^{-1}\boldsymbol{T}^{-\Ht}\boldsymbol{\Psi}_\mathrm{p}^\mathrm{H} \boldsymbol{r}_\mathrm{p}\notag\\
    =&\boldsymbol{T}(\boldsymbol{\Psi}_\mathrm{p}^\mathrm{H}\boldsymbol{\Psi}_\mathrm{p})^{-1}\boldsymbol{\Psi}_\mathrm{p}^\mathrm{H} \boldsymbol{r}_\mathrm{p}\notag\\
    =&\boldsymbol{T}\hat{\boldsymbol{\theta}}.
\end{align}
Therefore, the reconstructed SI satisfies
\begin{align}
    \boldsymbol{\Phi}_\mathrm{d}\hat{\boldsymbol{w}}=\boldsymbol{\Psi}_\mathrm{d}\hat{\boldsymbol{\theta}}.
\end{align}

\section*{Appendix B\\Proof of \textbf{Theorem 2}}
The estimation error of $\boldsymbol{\theta}$ under LS estimation can be written as
\begin{align}
    \hat{\boldsymbol{\theta}}-\boldsymbol{\theta}=\boldsymbol{\Psi}_{\ps}^\dagger(\epsilon_{\ps}+\boldsymbol{z}_{\ps}),
\end{align}
where the bias can be expressed as
\begin{align}\label{Estw_e}
    \boldsymbol{\Psi}_{\ps}^\dagger\epsilon_{\ps}=&\boldsymbol{G}_\mathrm{p}^{-1}\boldsymbol{\Psi}_\mathrm{p}^\mathrm{H}\boldsymbol{\epsilon}_\mathrm{p} \notag\\
    = &(\frac{1}{N_\mathrm{p}}\boldsymbol{G}_\mathrm{p})^{-1}\frac{1}{N_\mathrm{p}}\boldsymbol{\Psi}_\mathrm{p}^\mathrm{H}\boldsymbol{\epsilon}_\mathrm{p}.
\end{align}
When $x_\mathrm{p}[n]\overset{i.i.d.}{\sim}\mathcal{CN}(0,1)$, according to the law of large numbers, we have
\begin{align}
    \frac{1}{N_\mathrm{p}}\boldsymbol{G}_\mathrm{p} &= \frac{1}{N_\mathrm{p}} \sum_{n}^{N_\mathrm{p}}
    \boldsymbol{\psi}^*\big(x_{\ps}[n]\big)\boldsymbol{\psi}^{\rm T}\big(x_{\ps}[n]\big)\notag\\
    &\xrightarrow[N_\mathrm{p}\rightarrow\infty]{a.s.}\E{\boldsymbol{\psi}^*\big(x_{\ps}[n]\big)\boldsymbol{\psi}^{\rm T}\big(x_{\ps}[n]\big)}=\boldsymbol{I}_{L_w},
\end{align}
and
\begin{align}
    \frac{1}{N_\mathrm{p}}\boldsymbol{\Psi}_\mathrm{p}^\mathrm{H}\boldsymbol{\epsilon}_\mathrm{p} &= \frac{1}{N_\mathrm{p}}\sum_{n}^{N_\mathrm{p}}
    \boldsymbol{\psi}^*\big(x_{\ps}[n]\big)\epsilon_\mathrm{p}[n]\notag\\
    &\xrightarrow[N_\mathrm{p}\rightarrow\infty]{a.s.}\mathbb{E}\bigg\{\boldsymbol{\psi}^*\big(x_{\ps}[n]\big)\epsilon_\mathrm{p}[n]\bigg\}=\boldsymbol{0}.
\end{align}
Therefore, the expected estimation error satisfies:
\begin{align}
    \mathbb{E}_z\big\{\hat{\boldsymbol{\theta}}-\boldsymbol{\theta}|\boldsymbol{x_{\ps}}\big\}\xrightarrow[N_\mathrm{p}\rightarrow\infty]{a.s.}\boldsymbol{0}.
\end{align}

\section*{Appendix C\\Proof of \textbf{Proposition 1}}
First, for $x_{\rm p}(n)=\sqrt{u_{n}}e^{j\varphi_n}$, where $u_{n}\overset{i.i.d.}{\sim}{\rm GPD}(\xi,1-\xi)$ and $\varphi_n\overset{i.i.d.}{\sim}{\rm Unif}[0,2\pi)$, when $\xi<\frac{1}{P}$, the $p$-order moment $\E{u_n^p}$ for $p\le P$ is finite, the expectation of $\boldsymbol{G}_{\rm p}$ thus can be expressed as
\begin{align}
    \mathbb{E}\{ \boldsymbol{G}_{\rm p}|\xi \}
    =&\mathbb{E}\Bigg\{ \sum_{n}^{N_{\rm p}}\boldsymbol{\psi}^*\big( x_{\rm p}[n]\big)\boldsymbol{\psi}^{\rm T}\big( x_{\rm p}[n]\big) \Bigg\}\notag\\
    =&N_{\rm p}\mathbb{E}\Big\{\boldsymbol{\psi}^*\big( x_{\rm p}[n]\big)\boldsymbol{\psi}^{\rm T}\big( x_{\rm p}[n]\big) \Big\}\notag\\
    =&N_{\rm p}\mathbb{E}\Big\{\boldsymbol{\psi}_P^*\big( x_{\rm p}[n]\big)\boldsymbol{\psi}_P^{\rm T}\big( x_{\rm p}[n]\big) \Big\}\otimes\boldsymbol{I}_{L}\notag\\
    =&N_{\rm p}\boldsymbol{R}_{\psi_P}(\xi)\otimes\boldsymbol{I}_{L}.
\end{align}
Particularly, when $\xi=0$, $\boldsymbol{R}_{\psi_P}(0)=\boldsymbol{I}_{(P+1)/2}$. Therefore, we have
\begin{align}
    \tr{\mathbb{E}\{ \boldsymbol{G}_{\rm p}|\xi \}^{-1}}=\frac{L}{N_{\ps}}\tr{\boldsymbol{R}_{\psi_P}^{-1}(\xi)}.
\end{align}

For ${\rm tr}(\boldsymbol{R}_{\psi_P}^{-1}(\xi))$, it can be derived that
\begin{align}
    \frac{{\ds}\ {\rm tr}(\boldsymbol{R}_{\psi_P}^{-1}(\xi))}{{\ds}\xi}\Bigg|_{\xi=0}=&-\tr{\boldsymbol{R}_{\psi_P}^{-1}(0)\boldsymbol{R}_{\psi_P}'(0)\boldsymbol{R}_{\psi_P}^{-1}(0)}\notag\\
    =&-\tr{\boldsymbol{R}_{\psi_P}'(0)}\notag\\
    =&-\frac{{\ds}}{{\ds}\xi}\E{\norm{\boldsymbol{\psi}_P(\sqrt{u_n})}^2}\bigg|_{\xi=0}\notag\\
    =&-\E{\norm{\boldsymbol{\psi}_P(\sqrt{u_n})}^2s(u|\xi)|\xi=0},
\end{align}
where the score function $s(u|\xi)$ of GPD with CDF $F_\xi(u)$ as:
\begin{align}
    F_\xi(u)=1-\Big(1+\frac{\xi u}{1-\xi}\Big)^{-1/\xi},
\end{align}
can be derived as
\begin{align}
    s(u|\xi=0)=&\frac{\partial}{\partial\xi}\log\left(\frac{\partial F_\xi(u)}{\partial u}\right)\Bigg|_{\xi=0}\notag\\
    =&\frac{\partial}{\partial\xi}\left(-u+\left(1-2u+\frac{u^2}{2}\right)\xi+O(\xi^2)
    \right)\Bigg|_{\xi=0}\notag\\
    =&1-2u+\frac{u^2}{2}.
\end{align}
Therefore, ${\rm tr}(\boldsymbol{R}_{\psi_P}'(0))$ can be calculated as
\begin{align}
    \tr{\boldsymbol{R}_{\psi_P}'(0)}=&\int_{0}^{\infty}\sum_{k=0}^{(P-1)/2}\frac{u(L_k^1(u))^2}{k+1}s(u|0)e^{-u}{\ds}u\notag\\
    =&\sum_{k=0}^{(P-1)/2}\left(1-2\cdot2(k+1)+\frac{1}{2}\cdot6(k+1)^2\right)\notag\\
    =&\frac{(P-1)(P+1)(P+2)}{8}>0,
\end{align}
for $P>1$, since
\begin{align}
    \int_{0}^{\infty}\frac{(L_k^1(u))^2}{k+1}u^a e^{-u}{\rm d}u
    =[\boldsymbol{J}^{a-1}]_{k+1,k+1}, 
\end{align}
where $a\in \mathbb{N}_+$, and the non-zero elements of the Laguerre recurrence matrix $\boldsymbol{J}$ satisfy
\begin{align}
    &[\boldsymbol{J}]_{i,i}=2i,\notag\\
    &[\boldsymbol{J}]_{i,i+1}=[\boldsymbol{J}]_{i+1,i}=-\sqrt{i(i+1)},
\end{align}
which is derived from the recursive relationship of the GLP for $k>0$ as follows: 
\begin{align}
    \frac{uL_{k}^1(u)}{\sqrt{k+1}}=&-\sqrt{(k+1)(k+2)}\frac{L_{k+1}^1(u)}{\sqrt{k+2}}+(2k+2)\frac{L_{k}^1(u)}{\sqrt{k+1}}\notag\\
    &-\sqrt{k(k+1)}\frac{L_{k-1}^1(u)}{\sqrt{k}}.
\end{align}
Therefore, it can be proven that
\begin{align}
    \exists0<\xi<\frac{1}{P}:\tr{\mathbb{E}\{ \boldsymbol{G}_{\rm p}|\xi \}^{-1}}<\tr{\mathbb{E}\{ \boldsymbol{G}_{\rm p}|0 \}^{-1}}
\end{align}

\section*{Appendix D\\Proof of \textbf{Proposition 2}}

First, for the pilot sequence $x_{\rm p}(n)=\sqrt{u_{n}}e^{j\varphi_n}$, $\boldsymbol{G}_{\ps}$ with $L=1$ can be rewritten as
\begin{align}
    \boldsymbol{G}_{\ps}(\boldsymbol{u})=&\sum_{n}^{L_{\ps}}\boldsymbol{\psi}_P(\sqrt{u_n})\boldsymbol{\psi}_P^{\Ht}(\sqrt{u_n})\notag\\
    =&\sum_{n}^{L_{\ps}}\boldsymbol{C}_P(u_n),
\end{align}
where $\boldsymbol{C}_P(u_n)=\boldsymbol{\psi}_P(\sqrt{u_n})\boldsymbol{\psi}_P^{\Ht}(\sqrt{u_n})$. For the reconstructed pilot sequence $x_{\rm p}(n;\delta)=\sqrt{u_{n,\delta}}e^{j\varphi_n}$, defining
\begin{align}
    u_{i,\delta}=u_i+\delta,\ u_{j,\delta}=u_j-\frac{\delta}{M},\ j\in S_0,
\end{align}
where $u_i=\max\limits_{n}u_n$, $S_0=\{ j: \norm{\boldsymbol{\psi}_P(\sqrt{u_j})}^2\in \mathcal{U}_0, u_j>0 \}$ and $M=|S_0|$, we have
\begin{align}
    J(\delta)=\tr{\boldsymbol{G}^{-1}_{\ps}(\boldsymbol{u}_\delta)}.
\end{align}
Using
\begin{align}
    \frac{d}{d\delta}
    \boldsymbol{G}_{\ps}^{-1}(\boldsymbol{u}_{\delta})
    =
    -
    \boldsymbol{G}_{\ps}^{-1}(\boldsymbol{u}_{\delta})
    \frac{{\ds}\boldsymbol{G}_{\ps}(\boldsymbol{u}_{\delta})}{{\ds}\delta}
    \boldsymbol{G}_{\ps}^{-1}(\boldsymbol{u}_{\delta}),
\end{align}
the derivative of $J(\delta)$ with respect to $\delta$ can be obtained by
\begin{align}
    J'(\delta)
    =
    -\tr{\boldsymbol{G}_{\ps}^{-1}(\boldsymbol{u}_\delta)
    \boldsymbol{G}_{\ps}'(\boldsymbol{u}_{\delta})
    \boldsymbol{G}_{\ps}^{-1}(\boldsymbol{u}_\delta)}.
\end{align}
Since
\begin{align}
    \frac{{\ds}\boldsymbol{G}_{\ps}(\boldsymbol{u}_{\delta})}{d\delta}\bigg|_{\delta=0}=\boldsymbol{C}_P'(u_i)-\sum_{j\in S_0}\frac{\boldsymbol{C}_P'(u_j)}{M},
\end{align}
defining
\begin{align}
    f_{\boldsymbol{u}}(\mu)=&\tr{\boldsymbol{G}_{\ps}^{-1}(\boldsymbol{u})\boldsymbol{C}_P(\mu)\boldsymbol{G}_{\ps}^{-1}(\boldsymbol{u})}\notag\\
    =&\boldsymbol{\psi}_P^{\Ht}(\sqrt{\mu})\boldsymbol{G}_{\ps}^{-2}(\boldsymbol{u})\boldsymbol{\psi}_P(\sqrt{\mu}),
\end{align}
the derivative of $J(\delta)$ at $\delta=0$ thus can be expressed as
\begin{align}
    \gamma=&
    {\rm tr}\Bigg(
    \boldsymbol{G}_{\ps}^{-1}(\boldsymbol{u})
    \Bigg(
    \sum_{j\in S_0}\frac{\boldsymbol{C}_P'(u_j)}{M}
    -
    \boldsymbol{C}_P'(u_i)
    \Bigg)
    \boldsymbol{G}_{\ps}^{-1}(\boldsymbol{u})
    \Bigg)\notag\\
    =&
    \sum_{j\in S_0}\frac{f'_{\boldsymbol{u}}(u_j)}{M}-f'_{\boldsymbol{u}}(u_i).
\end{align}
Therefore, the sufficient condition for the proposition
\begin{align}
    \exists 0<\delta<M\min_{j\in S_0}u_j:\ \tr{\boldsymbol{G}_{\ps}^{-1}(\boldsymbol{u_{\delta}})}<\tr{\boldsymbol{G}_{\ps}^{-1}(\boldsymbol{u})},
\end{align}
can be formulated as
\begin{align}
    \gamma=\sum_{j\in S_0}\frac{f'_{\boldsymbol{u}}(u_j)}{M}-f'_{\boldsymbol{u}}(u_i)<0.
\end{align}
Furthermore, with the Rayleigh quotient inequality, $f'_{\boldsymbol{u}}(\mu)$ can be bounded as
\begin{align}
    f'_{\boldsymbol{u}}(\mu)=&
    2\Re\left\{
    \boldsymbol{\psi}_P^{\Ht}(\sqrt{\mu})\boldsymbol{G}_{\ps}^{-2}(\boldsymbol{u})\frac{{\ds}\boldsymbol{\psi}_P(\sqrt{\mu})}{{\ds}\mu}
    \right\}\notag\\
    \le&
    \frac{2\norm{\boldsymbol{\psi}_P(\sqrt{\mu})}\norm{\dot{\boldsymbol{\psi}}_P(\sqrt{\mu})}}{\lambda_{\min}^2(\boldsymbol{G}_{\ps}(\boldsymbol{u}))},
\end{align}
where $\dot{\boldsymbol{\psi}}_P(\sqrt{\mu})={\ds}\boldsymbol{\psi}_P(\sqrt{\mu})/{\ds}\mu$. Moreover, since
\begin{align}
    f_{\boldsymbol{u}}(\mu)=&|l_{P,P}|^2[\boldsymbol{G}_{\ps}^{-2}(\boldsymbol{u})]_{\frac{P+1}{2},\frac{P+1}{2}}\mu^P+O(\mu^{P-1}),
\end{align}
if $u_i$ is sufficiently large, we have
\begin{align}
    f'_{\boldsymbol{u}}(u_i)=&P|l_{P,P}|^2[\boldsymbol{G}_{\ps}^{-2}(\boldsymbol{u})]_{\frac{P+1}{2},\frac{P+1}{2}}u_i^{P-1}+O(u_i^{P-2})\notag\\
    \ge&
    \eta_0\lambda_{\min}(\boldsymbol{G}_{\ps}^{-2}(\boldsymbol{u}))u_i^{P-1}\notag\\
    =&\frac{\eta_0}{\lambda^2_{\max}(\boldsymbol{G}_{\ps}(\boldsymbol{u}))}u_i^{P-1}.
\end{align}
Therefore, a more intuitive sufficient condition can be formulated as:
\begin{align}
    u_i^{P-1}>\beta_0\kappa^2(\boldsymbol{G}_{\ps}(\boldsymbol{u})),
\end{align}
where
\begin{align}
        \beta_0=\sup \left\{
        \frac{2}{\eta_0}\|\dot{\boldsymbol{\psi}}_P(\sqrt{u_j})\|\norm{\boldsymbol{\psi}_P(\sqrt{u_j})} \big|\ j\in S_0 
        \right\}.
\end{align}

}

 
\bibliographystyle{IEEEtran}
\bibliography{reference}

@inproceedings{bharadia2013full,
  title={Full duplex radios},
  author={Bharadia, Dinesh and McMilin, Emily and Katti, Sachin},
  booktitle={Proceedings of the ACM SIGCOMM 2013 conference on SIGCOMM},
  pages={375--386},
  year={2013}
}

@article{Brink24,
  title={Optimized sequences for nonlinearity estimation and self-interference cancellation},
  author={Fuchs, Ephraim and Handte, Thomas and Verenzuela, Daniel and ten Brink, Stephan},
  journal={IEEE Transactions on Communications},
  volume={73},
  number={2},
  pages={846--858},
  year={2024},
  publisher={IEEE}
}

@inproceedings{anttila2014modeling,
  title={Modeling and efficient cancellation of nonlinear self-interference in {MIMO} full-duplex transceivers},
  author={Anttila, Lauri and Korpi, Dani and Antonio-Rodr{\'\i}guez, Emilio and Wichman, Risto and Valkama, Mikko},
  booktitle={2014 IEEE Globecom Workshops (GC Wkshps)},
  pages={777--783},
  year={2014},
  organization={IEEE}
}

@inproceedings{anttila2013cancellation,
  title={Cancellation of power amplifier induced nonlinear self-interference in full-duplex transceivers},
  author={Anttila, Lauri and Korpi, Dani and Syrj{\"a}l{\"a}, Ville and Valkama, Mikko},
  booktitle={2013 Asilomar Conference on Signals, Systems and Computers},
  pages={1193--1198},
  year={2013},
  organization={IEEE}
}

@inproceedings{korpi2015adaptive,
  title={Adaptive nonlinear digital self-interference cancellation for mobile inband full-duplex radio: Algorithms and {RF} measurements},
  author={Korpi, Dani and Choi, Yang-Seok and Huusari, Timo and Anttila, Lauri and Talwar, Shilpa and Valkama, Mikko},
  booktitle={2015 IEEE Global Communications Conference (GLOBECOM)},
  pages={1--7},
  year={2015},
  organization={IEEE}
}

@article{ahmed2015all,
  title={All-digital self-interference cancellation technique for full-duplex systems},
  author={Ahmed, Elsayed and Eltawil, Ahmed M},
  journal={IEEE Transactions on Wireless Communications},
  volume={14},
  number={7},
  pages={3519--3532},
  year={2015},
  publisher={IEEE}
}

@article{korpi2016full,
  title={Full-duplex mobile device: Pushing the limits},
  author={Korpi, Dani and Tamminen, Joose and Turunen, Matias and Huusari, Timo and Choi, Yang-Seok and Anttila, Lauri and Talwar, Shilpa and Valkama, Mikko},
  journal={IEEE Communications Magazine},
  volume={54},
  number={9},
  pages={80--87},
  year={2016},
  publisher={IEEE}
}

@article{komatsu2021theoretical,
  title={Theoretical analysis of in-band full-duplex radios with parallel {H}ammerstein self-interference cancellers},
  author={Komatsu, Kazuki and Miyaji, Yuichi and Uehara, Hideyuki},
  journal={IEEE Transactions on Wireless Communications},
  volume={20},
  number={10},
  pages={6772--6786},
  year={2021},
  publisher={IEEE}
}

@article{komatsu2020iterative,
  title={Iterative nonlinear self-interference cancellation for in-band full-duplex wireless communications under mixer imbalance and amplifier nonlinearity},
  author={Komatsu, Kazuki and Miyaji, Yuichi and Uehara, Hideyuki},
  journal={IEEE Transactions on Wireless Communications},
  volume={19},
  number={7},
  pages={4424--4438},
  year={2020},
  publisher={IEEE}
}

@article{korpi2014full,
  title={Full-duplex transceiver system calculations: Analysis of {ADC} and linearity challenges},
  author={Korpi, Dani and Riihonen, Taneli and Syrj{\"a}l{\"a}, Ville and Anttila, Lauri and Valkama, Mikko and Wichman, Risto},
  journal={IEEE Transactions on Wireless Communications},
  volume={13},
  number={7},
  pages={3821--3836},
  year={2014},
  publisher={IEEE}
}

@article{mohammadi2023comprehensive,
  title={A comprehensive survey on full-duplex communication: Current solutions, future trends, and open issues},
  author={Mohammadi, Mohammadali and Mobini, Zahra and Galappaththige, Diluka and Tellambura, Chintha},
  journal={IEEE Communications Surveys \& Tutorials},
  volume={25},
  number={4},
  pages={2190--2244},
  year={2023},
  publisher={IEEE}
}

@article{smida2023full,
  title={Full-duplex wireless for 6{G}: Progress brings new opportunities and challenges},
  author={Smida, Besma and Sabharwal, Ashutosh and Fodor, G{\'a}bor and Alexandropoulos, George C and Suraweera, Himal A and Chae, Chan-Byoung},
  journal={IEEE Journal on Selected Areas in Communications},
  volume={41},
  number={9},
  pages={2729--2750},
  year={2023},
  publisher={IEEE}
}

@inproceedings{islam2019comprehensive,
  title={A comprehensive self-interference model for single-antenna full-duplex communication systems},
  author={Islam, Md Atiqul and Smida, Besma},
  booktitle={ICC 2019-2019 IEEE International Conference on Communications (ICC)},
  pages={1--7},
  year={2019},
  organization={IEEE}
}

@article{elsayed2020low,
  title={Low complexity neural network structures for self-interference cancellation in full-duplex radio},
  author={Elsayed, Mohamed and El-Banna, Ahmad A Aziz and Dobre, Octavia A and Shiu, Wanyi and Wang, Peiwei},
  journal={IEEE Communications Letters},
  volume={25},
  number={1},
  pages={181--185},
  year={2020},
  publisher={IEEE}
}

@inproceedings{ahmed2013self,
  title={Self-interference cancellation with nonlinear distortion suppression for full-duplex systems},
  author={Ahmed, Elsayed and Eltawil, Ahmed M and Sabharwal, Ashutosh},
  booktitle={2013 Asilomar Conference on Signals, Systems and Computers},
  pages={1199--1203},
  year={2013},
  organization={IEEE}
}

@article{li2018augmented,
  title={An augmented nonlinear {LMS} for digital self-interference cancellation in full-duplex direct-conversion transceivers},
  author={Li, Zhe and Xia, Yili and Pei, Wenjiang and Wang, Kai and Mandic, Danilo P},
  journal={IEEE Transactions on Signal Processing},
  volume={66},
  number={15},
  pages={4065--4078},
  year={2018},
  publisher={IEEE}
}

@ARTICLE{liao2025analog,
  author={Liao, Limin and Sun, Jun and Wang, Junzhi and Liu, Yingzhuang},
  journal={IEEE Transactions on Communications}, 
  title={Analog Self-Interference Cancellation in Full-Duplex Radios: A Fundamental Limit Perspective}, 
  year={2025},
  volume={73},
  number={12},
  pages={15220-15232},
  }

 




\vfill

\end{document}